\documentclass[prb,twocolumn,showpacs,a4paper]{revtex4}
\usepackage{amsmath}
\usepackage{amsfonts}
\usepackage{amssymb}
\usepackage{hyperref}
\usepackage{epsfig}
\usepackage{epstopdf}
\usepackage{graphicx}
\usepackage{enumerate}
\usepackage{multirow}
\usepackage{verbatim}
\usepackage{color}

\begin{document}

\title{Role of oxygen-oxygen hopping in the three-band copper-oxide model: quasiparticle weight, metal insulator and magnetic phase boundaries, gap values and optical conductivity }
\author{Xin Wang$^1$, Luca de' Medici$^2$, and Andrew J. Millis$^1$}
\affiliation{$^1$Department of Physics, Columbia University, 538
West 120$^{th}$ Street, New York, New York 10027, USA\\
$^2$Laboratoire de Physique des Solides, Universit\'e Paris-Sud,
CNRS, UMR 8502, F-91405 Orsay Cedex, France}
\date{\today}
\pacs{74.72.-h, 74.25.Gz, 71.10.Fd, 71.27.+a}

\begin{abstract}
We investigate the effect of oxygen-oxygen hopping on the three-band copper-oxide model relevant to high-$T_c$ cuprates, finding that the physics is changed only slightly as the oxygen-oxygen hopping is varied. The location of the metal-insulator phase boundary in the plane of interaction strength and charge-transfer energy shifts by  $\sim 0.5$ eV or less along the charge-transfer axis, the quasiparticle weight has approximately the same magnitude and doping dependence and the qualitative  characteristics of the electron-doped and hole-doped sides of the phase diagram do not change. The results confirm the identification of La$_2$CuO$_4$ as a material with an intermediate correlation strength. However, the  magnetic phase boundary as well as higher-energy features of the optical spectrum are found to depend on the  magnitude of the oxygen-oxygen hopping. We compare our results to previously published one-band and three-band model calculations.
\end{abstract}

\maketitle

\section{Introduction}
The basic structural unit of the copper-oxide high-temperature superconductors is the CuO$_2$ plane and the important electronic states are hybridized combinations of  Cu $3d$  and O $2p_\sigma$ orbitals. The relevant electronic configurations are variations around  the Cu $3d^9$/O $2p^6$ configuration expected in a simple ionic picture.   While the Cu $d^{10}$ configuration is relatively close to the  chemical potential, strong local correlations on the  Cu  mean that the Cu  $d^8$ state  is very far removed in energy, and this strongly affects the physics.\cite{Eskes90} (Correlations on the oxygen have also been argued to be important but in the cuprates the probability of having two holes on the same oxygen is apparently  small enough that these may be neglected.)   The minimal model encoding this electronic structure  is the ``three-band'' model \cite{Mattheiss87, Emery87, Varma87, Andersen95} (defined more precisely below)  which involves the Cu $3d_{x^2-y^2}$ and O $2p_\sigma$ orbitals, along with local correlations on the Cu site and Cu-O and O-O hybridization.  While a qualitative understanding of the model was obtained early on \cite{Zaanen85, Kim89, Grilli90, Kotliar91,Dopf92}, modern theoretical methods, in particular Dynamical Mean Field Theory (DMFT), \cite{Georges96,Kotliar06} have added considerably to our understanding of the electronic structure of correlated electron materials. In particular, single-site DMFT predicts that at a carrier concentration of one hole per unit cell the three-band model undergoes a paramagnetic metal to paramagnetic insulator transition if the charge-transfer energy is decreased below a critical value. This critical value defines a correlation strength: one may characterize a material as being strongly correlated if the charge-transfer energy is such that the paramagnetic insulator phase is obtained, and having weak to intermediate correlation if not. Locating actual materials on this continuum of interaction strength is of interest.

Dynamical mean-field theory  has been used by several groups to study a simplified version of the three-band model, in which the oxygen-oxygen hopping is neglected. The pioneering single-site DMFT work of Georges {\sl et al.}\cite{Georges93} and Z\"olfl {\sl et al.}\cite{Zolfl98} has been followed in more recent years by cluster dynamical mean-field studies by Macridin {\sl et al.}\cite{Macridin05} and by single-site DMFT studies.\cite{Craco09}  Recent work has argued that in modeling specific materials such as La$_2$CuO$_4$ it is important to use parameters obtained from band theory calculations\cite{Weber08,Weber10a,Weber10b} and has stressed, in particular, the importance of including oxygen-oxygen hopping. 

Each of these papers considered one particular set of parameter values; however, for a comprehensive understanding of the physics and because it is not clear that any one method determines parameters accurately, it is important to determine how the physics changes as parameters are varied. 

In a previous paper\cite{demedici09} we presented a comprehensive study of  a version of the three-band model in which direct oxygen-oxygen hopping was neglected. While this study provided a qualitatively reasonable account of some aspects of cuprate physics, some aspects of the model were found to be in disagreement with experimental data, including the particle-hole asymmetry in the magnetic phase boundary, the optical absorption strength in the region just above the insulating gap, and the dependence on doping of the high energy optical absorption. References~\onlinecite{Weber08,Weber10a,Weber10b} presented results that differed in some respects from those in our work\cite{demedici09} and argued that the differences arose in part from the use of a more realistic band structure, including, in particular, oxygen-oxygen hopping.  

In this paper we extend our previous study to consider the consequences of oxygen-oxygen hopping. We find that inclusion of oxygen-oxygen hopping at values which are physically reasonable and consistent with band theory calculations does not appreciably change the metal to charge-transfer insulator phase diagram or the principal  characteristics of the electron- and hole-doped states, nor does it resolve the  contradiction with data in the magnitude of the above-gap conductivity; however, inclusion of oxygen-oxygen hopping does resolve the difficulties with the magnetic phase boundary and the very high energy conductivity. Our results indicate  that  differences between our results and those of Refs.~\onlinecite{Weber08,Weber10a,Weber10b} arise mainly from differences not in band structure but in correlation parameters ($U$ and charge-transfer gap) and, in some cases, from different results calculated for the same parameters.

The remainder of this paper is organized as follows: Sec.~\ref{formalism} presents the formalism, describing the models we studied and methods that we used. In Sec.~\ref{results} we present numerical results for the phase diagram, staggered magnetization and spectral functions. In Sec.~\ref{optics} we present the optical conductivities calculated from the three-band model and compare them to the conductivities calculated for  the one-band model. We compare our result to the result in Ref.~\onlinecite{Weber08} in Sec.~\ref{comparison}. Sec.~\ref{conclusion} is the conclusion.

\section{Formalism}\label{formalism}

\begin{figure}[]
    \centering
    \includegraphics[width=0.8\columnwidth]{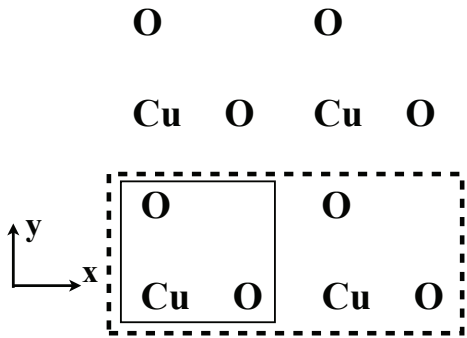}
    \caption{Sketch of a portion of the CuO$_2$ plane, with the coordinate and choice of unit cell in the paramagnetic (solid line) and antiferromagnetic (dashed line) cases indicated.}
    \label{geometry}
\end{figure}

The important structural unit for high-$T_c$ superconductivity is the CuO$_2$ plane, which in its idealized form is a square planar array of CuO$_2$ units. A small portion of the plane  is sketched in Fig.~\ref{geometry}.  We adopt the coordinate system shown in Fig.~\ref{geometry} and measure lengths in units of the nearest neighbour Cu-Cu distance so that the two lattice vectors are ${\boldsymbol{\hat x}}$ and  ${\boldsymbol{\hat y}}$. Our choice of unit cell in the symmetry-unbroken phase is indicated by the CuO$_2$ unit enclosed in the solid box. We will also have occasion to consider a two sub-lattice antiferromagnetic (AFM) state with lattice vectors ${\boldsymbol{\hat u}}={\boldsymbol{\hat x}}+{\boldsymbol{\hat y}}$ and ${\boldsymbol{\hat v}}={\boldsymbol{\hat x}}-{\boldsymbol{\hat y}}$ of length $\sqrt{2}$,  for which we choose the unit cell enclosed by the dashed box. 

We assume that the electronic physics of the CuO$_2$ plane is described by the ``three-band'' model introduced in the high-$T_c$ context by Emery.\cite{Emery87} This model involves a $3d_{x^2-y^2}$ orbital centered on the Cu site and created by an operator $d^\dagger$, and two oxygen $2p_\sigma$ orbitals,  ${p_x}$ on the O displaced in the $x$-direction from the Cu and ${p_y}$  on the O displaced in the $y$-direction. The creation operators for these states are $p^\dagger_{x,y}$ respectively. In the AFM phase it is necessary to distinguish the two sub-lattices. 

The Hamiltonian $H=H_{\rm band}+H_{\rm int}$  includes a band theoretical part $H_{\rm band}$  expressing the possibility of electron hopping from $d$ to $p$ orbitals and from $p$ to $p$ orbitals, and an interaction term. Following common practice we take the interaction part to be a repulsive interaction on the Cu d-sites (labeled by $j$)
\begin{equation}
 H_{\rm int}=U\sum_jn_{d\uparrow,j}n_{d\downarrow,j}.\label{Hint}
\end{equation}
with $U\sim 8-9$eV.\cite{Mila88,Veenendal94} Because of the low probability that any oxygen site has two holes we neglect any interaction on the O-sites. 

We write the band theoretic part of the Hamiltonian in momentum space. We adopt the Fourier transform convention
\begin{equation}
c_j^\alpha=\frac{1}{N}\sum_{\boldsymbol{k}}e^{i{\boldsymbol{k}}\cdot \left({\boldsymbol{R}}_j+{\boldsymbol{\rho}}^\alpha\right)}c^\alpha_{\boldsymbol{k}}
\label{FTconvention}
\end{equation}
with $\boldsymbol{R}_j$ the position of unit cell $j$, $\boldsymbol{\rho}^\alpha$ the position of atom $\alpha$ within the unit cell and $N$ the number of states in the crystal.  

For the paramagnetic (PM) phase we order the basis as $|\psi\rangle=\left({d_{\boldsymbol{k}\sigma}},{p_{x,\boldsymbol{k}\sigma}},{p_{y,\boldsymbol{k}\sigma}}\right)$, with $\boldsymbol{k}$ a wavevector in the full two-dimensional Brillouin zone of the problem and $\sigma$ spin indices, and obtain the band theory part of the Hamiltonian as a tight-binding model with  a$3\times3$ matrix:\cite{Andersen95,Wang10,Loewdin51}
\begin{widetext}
\begin{equation}
{\bf H}^{\rm PM}=\left(\begin{array}{ccc}
\varepsilon_d & 2it_{pd}\sin\frac{k_x}{2} & 2it_{pd}\sin\frac{k_y}{2}\\
-2it_{pd}\sin\frac{k_x}{2} & \varepsilon_p+2t_{pp}(\cos k_x-1) & 4t_{pp}\sin\frac{k_x}{2}\sin\frac{k_y}{2}\\
-2it_{pd}\sin\frac{k_y}{2} &
4t_{pp}\sin\frac{k_x}{2}\sin\frac{k_y}{2} &
\varepsilon_p+2t_{pp}(\cos k_y-1)
\end{array}\right)\label{Hpm}
\end{equation}
\end{widetext}

Here  $t_{pd}$ is the copper-oxygen hopping, $t_{pp}$ is the oxygen-oxygen hopping. It will be helpful in later discussions to define the $p$-$d$ level splitting $\Delta=\varepsilon_p-\varepsilon_d$, which we also vary.

Different authors have obtained different forms and parameters for the downfolded Hamiltonian. For most of the paper we use the particular form of oxygen-oxygen hopping  obtained by O. K. Andersen and collaborators.\cite{Andersen95}  We take $t_{pd}=1.6$ eV and study different values of  $t_{pp}$, focusing mostly on $t_{pp}=0$ or $1.1$ eV\cite{Andersen95,Wang10} which bracket the range of values proposed in the literature.\cite{Weber10b,Hybertsen89,Hybertsen90,McMahan90,Korshunov05,Hanke10} 
We show that most of  the low energy physics is not sensitive to the precise value of the oxygen-oxygen hopping, while the magnetic phase boundary and higher energy optical conductivity do depend on this parameter. An alternative form for the oxygen-oxygen hopping, with the terms proportional to $t_{pp}$ on the diagonal of $H$ being absent, has been used by some authors.\cite{Hybertsen89,Mila88b,Korshunov05,Weber08,Weber10a,Weber10b,Hanke10} In Sec.~\ref{comparison} we use the Hamiltonian form presented by those authors, and the value $t_{pd}=1.41$ eV proposed by Refs.~\onlinecite{Weber08,Weber10a,Weber10b}.


To treat the  Neel AFM phase occurring at and near half-filling we divide the lattice into two sub-lattices $A$ and $B$, with the $A$ sub-lattice being that part shown within the solid box and the $B$ sub-lattice being the part remaining inside the dashed box in Fig.~\ref{geometry}. We Fourier transform using Eq.~\eqref{FTconvention} with appropriate intra-unit cell coordinates tied to the magnetic unit cell.   In  the basis $|\psi\rangle=\left({d^A_{\boldsymbol{k}\sigma}}, {p^A_{x,\boldsymbol{k}\sigma}}, {p^A_{y,\boldsymbol{k}\sigma}}, {d^B_{\boldsymbol{k}\sigma}}, {p^B_{x,\boldsymbol{k}\sigma}}, {p^B_{y,\boldsymbol{k}\sigma}}\right)$,  the band theoretic part of the Hamiltonian becomes  a $6\times 6$ matrix that we express in block form as
\begin{equation}
{\bf H}^{\rm AFM}=\left(\begin{array}{cc}
{\bf H}_A & {\bf H}_M\\
{\bf H}_M & {\bf H}_B
\end{array}\right).
\end{equation}
The $3\times3$ matrices are
\begin{equation}
\begin{split}
&{\bf H}_A={\bf H}_B\\&=\left(\begin{array}{ccc}
\varepsilon_d & t_{pd}e^{i\frac{k_x}{2}} & t_{pd}e^{i\frac{k_y}{2}}\\
t_{pd}e^{-i\frac{k_x}{2}} & \varepsilon_p-2t_{pp} &
2t_{pp}\cos\frac{k_x-k_y}{2}\\
t_{pd}e^{-i\frac{k_y}{2}} & 2t_{pp}\cos\frac{k_x-k_y}{2} &
\varepsilon_p-2t_{pp}
\end{array}\right),
\end{split}\label{HafmAB}
\end{equation}
\begin{equation}
{\bf H}_M=\left(\begin{array}{ccc}
0 & -t_{pd}e^{-i\frac{k_x}{2}} & -t_{pd}e^{-i\frac{k_y}{2}}\\
-t_{pd}e^{i\frac{k_x}{2}} & 2t_{pp}\cos k_x & -2t_{pp}\cos\frac{k_x+k_y}{2}\\
-t_{pd}e^{i\frac{k_y}{2}} & -2t_{pp}\cos\frac{k_x+k_y}{2} &
2t_{pp}\cos k_y
\end{array}\right),\label{HafmM}
\end{equation}

Later in the paper we will require the current operator ${\bf j}(\boldsymbol{k})$ which is a $3\times 3$ or $6\times6$ matrix in the PM or AFM cases respectively. Our Fourier transform convention, Eq.~\eqref{FTconvention} implies that  ${\bf j}=\delta {\bf H}/\delta k_x$ for both PM and AFM cases.  Tomczak and Biermann\cite{Tomczak09} noted that care is required in constructing the current operator: if a different Fourier transform convention is used, for example, omitting the $\boldsymbol{\rho}^\alpha$ terms in Eq.~\eqref{FTconvention},  additional terms in the current operator must be introduced to recover the correct result. These terms are not needed with the choice of convention used here.

We solve the model using the single-site DMFT\cite{Georges96, Kotliar06} primarily in conjunction with the continuous-time quantum Monte Carlo impurity solver in its hybridization-expansion (CT-HYB) form.\cite{Werner06,Gullrev} The specifics are described in Refs.~\onlinecite{demedici09,Wang10}. For the phase diagram scan (Fig.~\ref{phase}) and the doping dependence of the quasiparticle
renormalization factor $Z=(1-\partial\Sigma/\partial\omega)^{-1}$ (Fig.~\ref{Zplot}) we used the zero temperature Exact Diagonalization(ED)
\cite{Caffarel94,Capone04} method. We also cross-checked the results of CT-HYB and ED calculations.

 CT-HYB is formulated in imaginary time and to obtain real-frequency information we perform analytic continuation of the imaginary-axis self-energies  using the method of Ref.~\onlinecite{Wang09}. From the self-energies we calculate the electron Green's function ${\bf G}$ at frequency  $z$ (we choose the zero of energy such that the chemical potential $\mu=0$)  as
\begin{equation}
{\bf G}(z, \boldsymbol{k})=\left[z{\bf 1}-{\bf H}-{\bf \Sigma}(z,\boldsymbol{k})\right]^{-1}
\label{Gdef}
\end{equation}
with ${\bf \Sigma}$ a matrix in which all entries vanish except the $d$-$d$ components.

The electron spectral function
${\mathbf A}$ is
\begin{equation}
{\bf A}(\omega,\boldsymbol{k})=
\frac{{\bf G}(\omega,\boldsymbol{k})-{\bf G}^\dagger(\omega,\boldsymbol{k})}{2i},\label{Adef}
\end{equation}

The optical conductivities are obtained from:\cite{Millis05}
\begin{align}
\sigma(\Omega)&=\frac{2e^2}{\hbar c}\int_{-\infty}^\infty \frac{d\omega}{\pi}\int
\frac{d^2\boldsymbol{k}}{(2\pi)^2}\frac{f(\omega)-f(\omega+\Omega)}{\Omega} \nonumber
\\
&\times{\mathrm{Tr}}\left[{\bf j}(\boldsymbol{k}){\bf A}(\omega+\Omega,\boldsymbol{k}){\bf
j}(\boldsymbol{k}){\bf A}(\omega,\boldsymbol{k})\right], \label{sigmamatrix}
\end{align}
where $c$ is the $c$-axis lattice constant, $f(\omega)$ is the Fermi function, the $\boldsymbol{k}$-integral is over
the magnetic Brillouin zone for the AFM case and the full zone in the PM case.

We will also consider the  integrated optical spectral weight:
\begin{equation}
K(\Omega)=\frac{2}{\pi}\int_0^\Omega \left(\frac{\hbar c}{e^2}\right)\sigma(\Omega')d\Omega'\label{Kdef}
\end{equation}
The $\Omega\rightarrow\infty$ limit, denoted $K\equiv K(\Omega\rightarrow\infty)$, yields the familiar $f$-sum rule:\cite{Millis04}
\begin{equation}
 K=2\int_{-\infty}^\infty \frac{d\omega}{\pi}\int
\frac{d^2\boldsymbol{k}}{(2\pi)^2}f(\omega)
\mathrm{Tr}\left[\frac{\partial^2{\bf
H}(\boldsymbol{k})}{\partial k_x^2}{\bf A}(\omega,\boldsymbol{k})\right]\label{sumrule}
\end{equation}

For comparison we have studied the one-band model described (in the AFM phase) by\cite{Andersen95, Lin09}
\begin{equation}
{\bf H}_{\rm 1band}^{\rm AFM}=\left(
\begin{array}{cc}
\varepsilon_d-4t'\cos k_x\cos k_y & -2t(\cos k_x+\cos k_y)\\
-2t(\cos k_x+\cos k_y) & \varepsilon_d-4t'\cos k_x\cos k_y
\end{array}
\right) \label{H1band}
\end{equation}
with
\begin{equation}
H_{\rm int}=U_{\rm eff}\sum_jn_{d\uparrow,j}n_{d\downarrow,j}.
\end{equation}
We take the conventional choice of parameters $t=0.37$ eV, $t'=-0.3t$, and
choose $U_{\rm eff}=9t$ to reproduce the correlation gap.

\section{Results: phase diagram and electron spectral function}\label{results}

\subsection{Paramagnetic phase diagram}

\begin{figure}[]
    \centering
    \includegraphics[width=6.5cm, angle=-90]{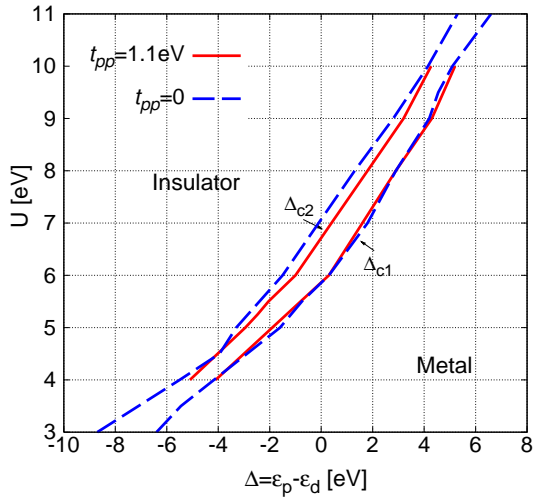}
    \caption{Metal-insulator phase diagram in plane of interaction $U$ and $p$-$d$ level splitting $\Delta$ for one hole per CuO$_2$ unit in the PM phase at zero temperature, using ED as the impurity solver. Solid line (red online): phase boundary calculated for $t_{pp}=1.1$ eV; dashed line (blue online): phase boundary calculated for $t_{pp}=0$.\cite{demedici09}}
    \label{phase}
\end{figure}

Fig.~\ref{phase} shows the metal-insulator phase diagram calculated at a carrier concentration of  one hole per CuO$_2$ unit in the PM phase. There are two important features of the phase diagram: $\Delta_{c1}$, the value of the charge-transfer energy at which the insulating solution loses stability and $\Delta_{c2}$ at which the $T=0$ metallic solution loses stability. The dashed line (blue online) shows the $t_{pp}=0$ case studied  previously\cite{demedici09}; the results presented here are generally  consistent with our previous work but the higher accuracy data available to us now have slightly altered the phase boundaries.  The solid line (red online) shows the phase boundary for  $t_{pp}=1.1$eV, confirming that the effect of oxygen-oxygen hopping on the  phase boundary is small, corresponding to about a $0.5$ eV shift in the critical $\Delta_{c2}$ and a  smaller shift in $\Delta_{c1}$. We note that as one turns on AFM order the part of metallic regime that immediately follows $\Delta_{c1}$ will have a gap, which we have shown in Ref.~\onlinecite{demedici09} to fit the experimentally observed value. We  focus on this regime in the following discussion.

\subsection{Magnetization}

\begin{figure}
    \centering
    \includegraphics[width=6.5cm, angle=-90]{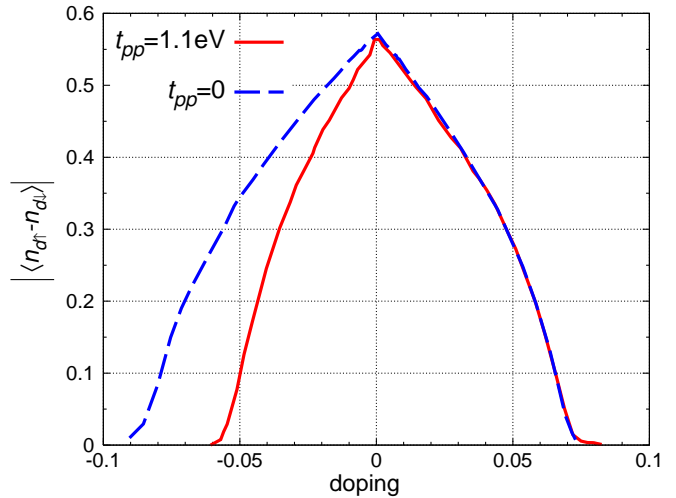}
    \caption{Staggered magnetization ($\langle m\rangle=\left|\left\langle n_{d\uparrow}-n_{d\downarrow}\right\rangle\right|$) v.s. doping in AFM phase. Solid line (red online): $t_{pp}=1.1$ eV, $\Delta=4.2$ eV; dashed line (blue online): $t_{pp}=0$, $\Delta=4.5$ eV. Positive doping values indicate electron doping and negative doping values indicate hole doping. Common parameters: $U=9$ eV, $t_{pd}=1.6$ eV, and
    $T=0.1$ eV.}
    \label{magn}
\end{figure}

Fig.~\ref{magn} shows the staggered magnetization
$\langle m\rangle=\left|\left\langle n_{d\uparrow}-n_{d\downarrow}\right\rangle\right|$
as a function of doping for the model (solved of course allowing for the possibility of antiferromagnetism) with $t_{pp}=0$ and $t_{pp}=1.1$ eV.  For each value of $t_{pp}$ we chose a $\Delta$ such that the undoped system has a gap of 1.8 eV in the AFM phase. The values of $\Delta$ for the two $t_{pp}$ are only slightly different: for $t_{pp}=1.1$ eV, $\Delta=4.2$ eV; and for $t_{pp}=0$, $\Delta=4.5$ eV. This is consistent with our discussion that $t_{pp}$ shows only a slight change on the metal/insulator phase diagram. It is important to note that the calculations are performed at a relatively high temperature $T\approx 1200$ K; the magnetization is not yet saturated. The dashed line (blue online) is the $t_{pp}=0$ result: we see a clear particle-hole asymmetry, with antiferromagnetism disappearing at about 0.09 hole doping and 0.07 electron doping. The solid line (red online) shows the $t_{pp}=1.1$ eV result: at the electron doped side it is almost identical with the $t_{pp}=0$ result, while at the hole doped side antiferromagnetism vanishes at about 0.06 hole doping. Thus oxygen-oxygen hopping has a strong effect on the hole-doped magnetic phase boundary. Comparison to experiment is complicated by the incommensurate (stripe) magnetism observed on the hole-doped side, and its complicated interplay with superconductivity and the pseudogap. None of these are accurately treated by the single-site DMFT approximation.

Our calculations reveal that the magnetic phase boundary is affected more strongly by changes in $t_{pp}$ than the metal-charge transfer insulator phase boundary. A possible reason may be revealed by the observation that oxygen-oxygen hopping affects the doping dependence of the magnetic phase boundary much more strongly than it affects the Neel temperature of the undoped phase.  As the doping is increased, the calculated Neel temperature drops, as does the size of the low-$T$ moment. Thus the magnetic transition becomes more weak-coupling-like at higher doping, and it is well known that weak coupling transitions are sensitive to details of the fermiology, which is in turn sensitive to the oxygen-oxygen hopping, whereas the stronger coupling phenomena are governed more by local physics, which is less sensitive to details.

\subsection{Spectral functions}

\begin{figure}
    \centering
    \includegraphics[width=6.5cm, angle=-90]{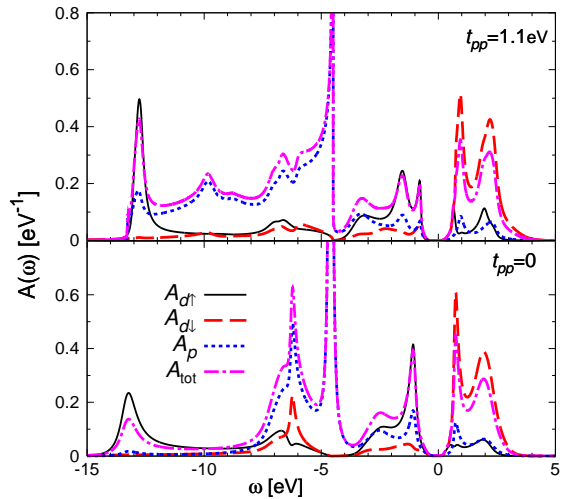}
    \caption{Spectral functions for undoped AFM case. $A_{d\uparrow}$ and $A_{d\downarrow}$ shows the majority and minority spin density of states for $d$-orbitals. $A_p$ is the total $p$ density of states for a given spin ($=A_{p_x}+A_{p_y}$) and $A_{\rm tot}$ shows spin averaged total density of states [$A_{\rm tot}=A_p+(A_{d\uparrow}+A_{d\downarrow})/2$]. Upper Panel: $t_{pp}=1.1$ eV, $\Delta=4.2$ eV, $\varepsilon_d=-8.7$ eV,
    $\varepsilon_p=-4.5$ eV;
    Lower panel: $t_{pp}=0$, $\Delta=4.5$ eV, $\varepsilon_d=-9.1$ eV, $\varepsilon_p=-4.6$ eV. Common parameters: $U=9$ eV, $t_{pd}=1.6$ eV, $T=0.1$ eV. Fermi energy is at zero.}
    \label{dosAFM}
\end{figure}

Fig.~\ref{dosAFM} shows the spectral functions computed for the undoped insulating AFM state. 
We choose parameters such that the material is on the metallic side of the metal to charge-transfer insulator phase diagram, so that the insulating gap arises (in the model) from the presence of AFM order. The spectral gap is determined by proximity to the phase boundary shown in Fig.~\ref{phase}.The small shift in the phase boundary with $t_{pp}$ indicated in Fig.~\ref{phase} would lead to a shift in the gap values computed for two different $t_{pp}$ values at the same $\Delta$ and $U$. Because the value of $\Delta$ cannot be determined \emph{a priori}, we study the effect of $t_{pp}$ on the spectral form at fixed physical excitation gap of about 1.8 eV, with $\Delta$ slightly adjusted correspondingly. With the value of $t_{pd}$ we have chosen here we find that if the model has a small enough $\Delta$ to be in the PM insulating phase then as AFM order is turned on the gap will be increased by $0.3$ eV or more (the increase is largest at $t_{pp}=0$ and decreases as $t_{pp}$ increases).

The bottom panel in Fig.~\ref{dosAFM} is the $t_{pp}=0$ case studied in Ref.~\onlinecite{demedici09}. We see a gap of about $1.8$ eV near the Fermi energy. The states just below the gap have an appreciable oxygen character and are identified with Zhang-Rice singlets while the states above the gap have almost an exclusively copper character and are identified with the upper Hubbard band. Examination of the $\boldsymbol{k}$-space structure shows that the gap is indirect in all cases, with valence band maximum at $\left(\pi/2, \pi/2\right)$ and conduction band maximum at $\left(0, \pi\right)$. The indirect nature of the gap was noted in Refs.~\onlinecite{Weber08,Weber10a,Weber10b}. The  non-bonding oxygen band is visible as a delta function centered at $\omega=\varepsilon_p=-4.6$ eV. The  peak at binding energy $\omega\sim 12$ eV represents the Cu-$d^8$ state, which is displaced below the $U=9$ eV by level repulsion with the oxygen.  Turning to the upper panel, the $t_{pp}=1.1$ eV case, we see that the height of the gap-edge peak in the Zhang-Rice region is reduced, and the Zhang-Rice absorption is spread out over a wider energy range. Also, the structure at higher binding energy is altered. The delta function at $\omega=\varepsilon_p=-4.6$ eV is replaced by an integrable singularity and the oxygen bands are broadened. The peak representing the  Cu-$d^8$ state is changed in form, but the $t_{pp}$ evidently has little effect on its energy.  For smaller $U$ or larger $t_{pp}$ the $d^8$  feature may be absorbed into the oxygen bands.  

\begin{figure}
    \centering
    \includegraphics[width=7cm, angle=-90]{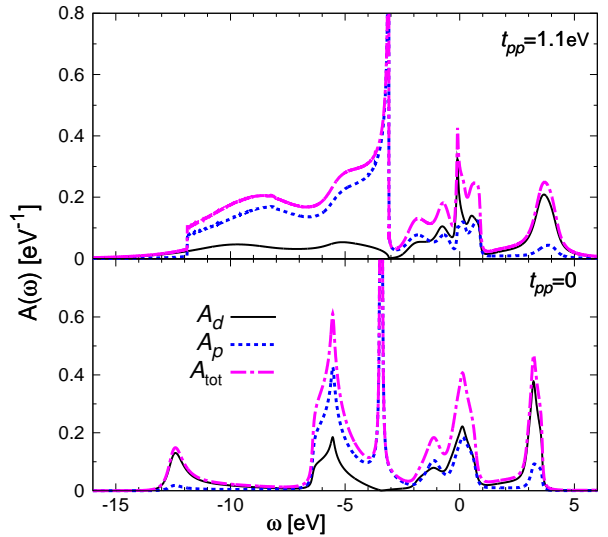}
    \caption{Spectral functions for 0.18 hole doped PM case. $A_d$, $A_p$ and $A_{\rm tot}$ shows respectively spin averaged $d$, $p$ and total density of states.
    $A_{\rm tot}=A_p+A_d$. Upper Panel: $t_{pp}=1.1$ eV, $\Delta=4.2$ eV, $\varepsilon_d=-7.29$ eV,
    $\varepsilon_p=-3.09$ eV;
    Lower panel: $t_{pp}=0$, $\Delta=4.5$ eV, $\varepsilon_d=-7.93$ eV, $\varepsilon_p=-3.43$ eV. Common parameters: $U=9$ eV, $t_{pd}=1.6$ eV, $T=0.1$ eV.}
    \label{dosPM1}
\end{figure}

Calculations of both the hole-doped (shown in Fig.~\ref{dosPM1}) and electron-doped cases (not shown here) show the same behaviour that a non-zero $t_{pp}$ expands the non-bonding oxygen band to a width of $8t_{pp}$, consistent with the above discussions.  We note that at $t_{pp}=1.1$ eV the bands change enough to absorb the $d^8$ peak into the oxygen continuum, while if the value of $t_{pp}$ is smaller, e.g. around 0.6 eV as in Refs.~\onlinecite{Hybertsen89,Hybertsen90,McMahan90}, the width of the oxygen band would be approximately one half of that of the result of $t_{pp}=1.1$ eV. Fig.~\ref{dosPM1} shows  that the Zhang-Rice singlet  is not affected in the non-zero $t_{pp}$ case, in the sense that the spectral weights of $d$-content and $p$-content remain approximately equal.  We have also found (not shown, but see the discussion of the conductivity below) that the self-energy at low Matsubara frequencies depends on $t_{pp}$ only at the $5\%$ level, if the distance from the metal-charge transfer insulator phase boundary is held constant.  Therefore we conclude that inclusion of $t_{pp}$ does not have an important effect on the characteristics of near-Fermi-surface states, in particular the Zhang-Rice physics. However it does affect the high binding energy features.

\section{Results: Optical conductivity}\label{optics}

\subsection{Results of the three-band model}

\begin{figure}
    \centering
(a)\includegraphics[width=6.5cm, angle=-90]{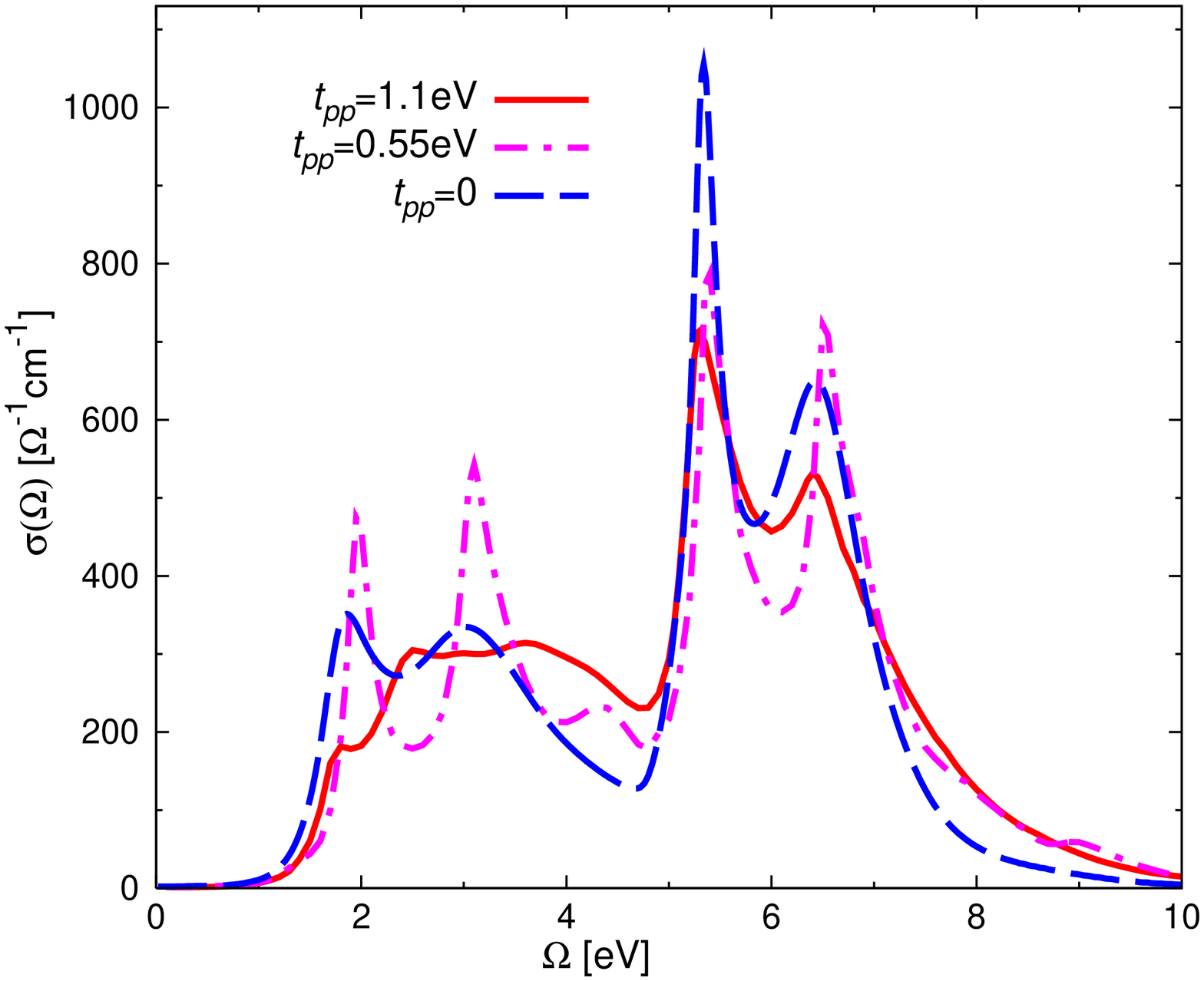}
(b)\includegraphics[width=6.5cm, angle=-90]{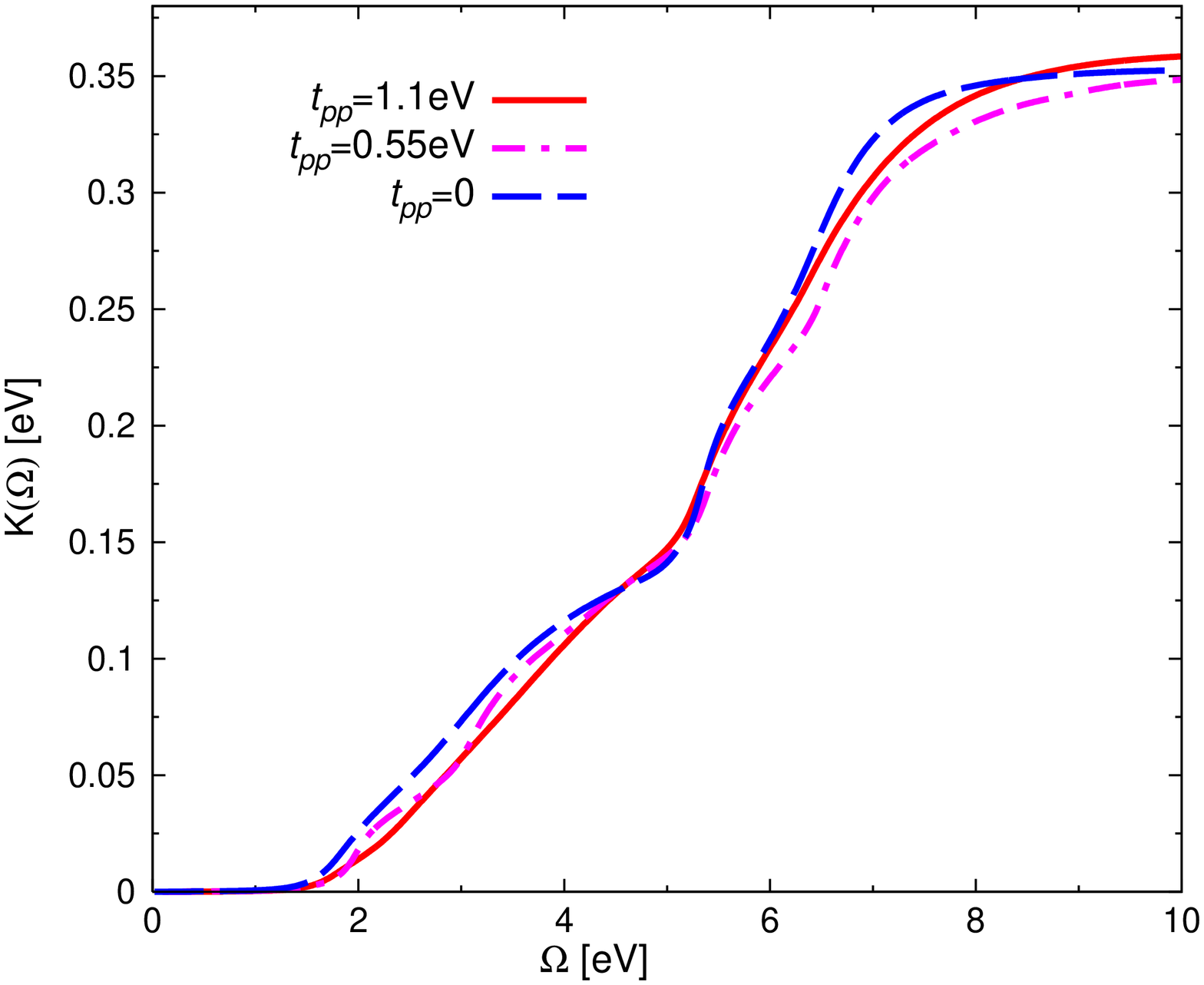}
    \caption{Optical conductivity [Panel (a)] and the corresponding integrated spectral weight [Panel (b)]
    in the undoped AFM case at various $t_{pp}$ values. For $t_{pp}=1.1$ eV and 0, the parameters are the same as in Fig.~\ref{dosAFM}. For $t_{pp}=0.55$ eV,
    the parameters are
    $\Delta=4.35$ eV, $\varepsilon_d=-9.9$ eV, $\varepsilon_p=-5.55$ eV, $U=9$ eV, $t_{pd}=1.6$ eV, $T=0.1$ eV.}
    \label{condAFM}
\end{figure}

Fig.~\ref{condAFM} shows the computed optical conductivity in the insulating AFM phase over a wide frequency range, along with the integrated spectral weight for three different values of $t_{pp}$. (The indirect nature of the gap in the single-particle spectrum means that the conductivity rises slowly above its onset.) $\Delta$ is adjusted such that the gap is around $1.8$ eV.  In the  frequency range  between 1.8 eV and 5 eV the conductivity arises from transitions between the upper Hubbard band and the Zhang-Rice singlet bands. The detailed line shape depends on $t_{pp}$, in a manner which is roughly consistent with the change in spectral function displayed in Fig.~\ref{dosAFM}.  The $t_{pp}=0$ and $0.55$ eV cases have a two-peak structure which is absent in the $t_{pp}=1.1$ eV calculation. The $t_{pp}=0.55$ eV conductivity is sharper than the other two. Some of the difference in the traces arise from uncertainties in the analytic continuation, but there is a clear trend with the integrated spectral weights. As  $t_{pp}$ is increased the weight in the near gap-edge region decreases, but the integrated areas reach similar values by $\Omega=5$ eV [cf. Fig.~\ref{condAFM}(b)]. We believe that the robustness of the evaluation of the gap and similarity of the integrated spectral weight means that the main features of our calculations are reliable.

The sharper peaks between 5 eV and 8 eV come from the transition between the upper Hubbard band and the mainly oxygen band at and below $\omega=\varepsilon_p$. All three curves show a two-peak structure in these range. The two-peak structure is a reflection of a similar structure visible  in the upper Hubbard band density of states. This structure has been previously discussed in Refs.~\onlinecite{Krivenko06, Wang09, Gull10} and the variation in peak height may be traced to the changes in the density of occupied states. The integrated spectral weights are similar in all curves and we have verified that the kinetic energy $K(\Omega)\rightarrow\infty$ is consistent with the value computed directly from the Matsubara axis.\cite{Lin09}  We see that $t_{pp}$ does not have significant effect on the conductivity in undoped AFM case.

\begin{figure}
    \centering
    \includegraphics[width=6.5cm, angle=-90]{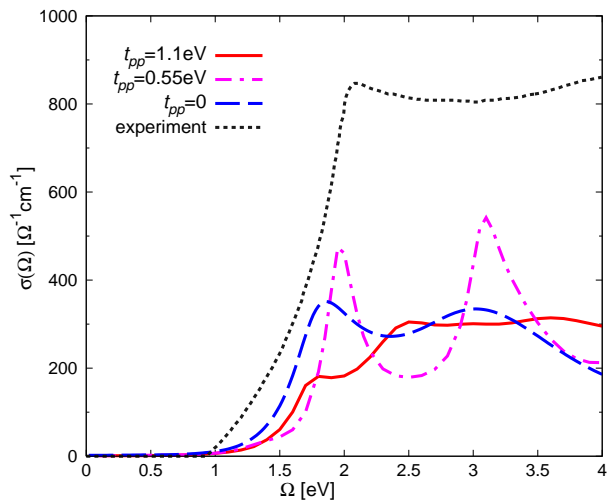}
    \caption{Low-frequency part of the optical conductivity
    in the undoped AFM case  at various $t_{pp}$ values. For $t_{pp}=1.1$ eV and 0, the parameters are the same as in Fig.~\ref{dosAFM}. For $t_{pp}=0.55$ eV,
    the parameters are
    $\Delta=4.35$ eV, $\varepsilon_d=-9.9$ eV, $\varepsilon_p=-5.55$ eV, $U=9$ eV, $t_{pd}=1.6$ eV, $T=0.1$ eV. The experimental curve is reproduced from Ref.~\onlinecite{Uchida91}.}
    \label{condAFMlowfreq}
\end{figure}

Fig.~\ref{condAFMlowfreq} shows an expanded view of the near gap region, along with the experimental conductivity.\cite{Uchida91} The magnitude of the conductivity, as noted in Ref.~\onlinecite{Weber08,demedici09} is still a factor of two smaller than experiments.\cite{Uchida91} Ref.~\onlinecite{Weber10b} states that inclusion of apical oxygen bands improves the comparision to experiments; the validity of the Peierls phase approximation at these energies is also open to question.  Further investigation of this issue is important.

\begin{figure}
    \centering
    \includegraphics[width=6.5cm, angle=-90]{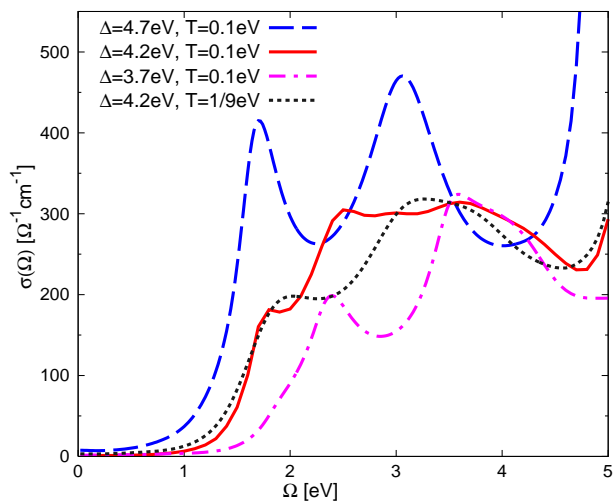}
    \caption{Low-frequency part of the optical conductivity
    in the undoped AFM case at different $\Delta$ values, with $t_{pp}$ fixed at 1.1eV and temperatures indicated. For $\Delta=4.2$ eV, the parameters are the same as in Fig.~\ref{dosAFM}. For $\Delta=4.7$ eV,  $\varepsilon_d=-8.7$ eV, $\varepsilon_p=-4.0$ eV; For $\Delta=3.7$ eV, $\varepsilon_d=-8.7$ eV, $\varepsilon_p=-5.0$ eV. Common parameters: $U=9$ eV, $t_{pd}=1.6$ eV. The values of staggered magnetization are 0.59 ($\Delta=4.7$ eV, $T=0.1$ eV), 0.56 ($\Delta=4.2$ eV, $T=0.1$ eV), 0.47 ($\Delta=3.7$ eV, $T=0.1$ eV) and 0.46 ($\Delta=4.2$ eV, $T=1/9$ eV) respectively.}
    \label{condAFMlowfreqdiffdelta}
\end{figure}

Fig.~\ref{condAFMlowfreqdiffdelta} shows the low-frequency part of the optical conductivity in the undoped AFM case at different $\Delta$ values (indicated on the figure), with $t_{pp}$ fixed at 1.1 eV. We see that as $\Delta$ is increased (pushing the system further into the PM metal region), the gap becomes smaller and the magnitude of the conductivity becomes larger; as $\Delta$ goes into the PM insulating regime, the gap becomes larger and the conductivity magnitude smaller.  We see that a $\Delta$ between 4.2 eV and 3.7 eV, i.e. slightly smaller than $\Delta_{c1}$ but noticeable larger than $\Delta_{c2}$ is required to place the gap in the experimentally observed range.   The calculations are done at fixed temperature ($T=0.1$ eV). Because the magnetization is not saturated, the different $\Delta$ values imply different staggered magnetizations. In general the Neel temperature is lower in the insulating regime than the metallic regime, therefore the magnetization $m$ for $\Delta=3.7$ eV ($m=0.47$) is smaller than that of $\Delta=4.7$ eV ($m=0.59$). To make sure that the different magnetization values do not change our conclusion, we tune the temperature for $\Delta=4.2$ eV calculation such that it has similar magnetization as the $\Delta=3.7$ eV one (shown as dotted line in Fig.~\ref{condAFMlowfreqdiffdelta}). We see that the gap essentially does not change, while the magnitude only changes slightly which may be attributed to the error of analytic continuation.

\begin{figure}
    \centering
    \includegraphics[width=6.5cm, angle=-90]{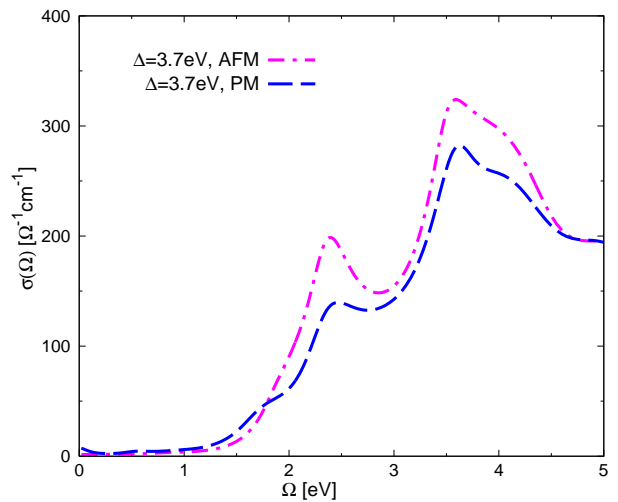}
    \caption{Comparison of the optical conductivity calculated at $\Delta=3.7$ eV (between $\Delta_{c1}$ and $\Delta_{c2}$) between AFM and PM cases. Parameters: $\varepsilon_d=-8.7$ eV, $\varepsilon_p=-5.0$ eV, $U=9$ eV, $t_{pd}=1.6$ eV, $t_{pp}=1.1$ eV, $T=0.1$ eV.}
    \label{condcompAFMPM}
\end{figure}

It is also interesting to consider the effect of antiferromagnetism on the gap magnitude. In studies of a one-band model with a bipartite lattice,\cite{Comanac08} a large effect was found, with the gap edge shifting up by $\sim40\%$ as the PM insulating state was allowed to become AFM. In our studies of the three-band model with $t_{pp}=0$\cite{demedici09} a non-negligible, but smaller shift was reported ($\sim0.6$ eV out of 3 eV). Other work \cite{Weber08} reported no observable shift. Fig.~\ref{condcompAFMPM} examines the issue for the case of $t_{pp}=1.1$ eV and $\Delta=3.7$ eV. Antiferromagnetism does correspond to an observable shift of the gap edge to higher frequency, but from Fig.~\ref{condcompAFMPM} one would conclude that the shift is smaller ($\sim0.2$ eV) than in the $t_{pp}=0$ case. However, a direct quantitative analysis of the gap implied by conductivity data such as is shown in Fig.~\ref{condcompAFMPM} is complicated by broadening arising from analytic continuation and high temperature. In our previous work\cite{Wang09} we advocated using a quasiparticle analysis, defining the gap edge from zeros of $\det[\omega{\bf 1}-{\bf \Sigma}-{\bf H}]$. Applying this method to the self-energies used in constructing the conductivities shown in Fig.~\ref{condcompAFMPM}  yields a gap of 1.5 eV in the PM insulating case and 2.0 eV in the AFM case.

\begin{figure}
    \centering
(a)\includegraphics[width=6.5cm, angle=-90]{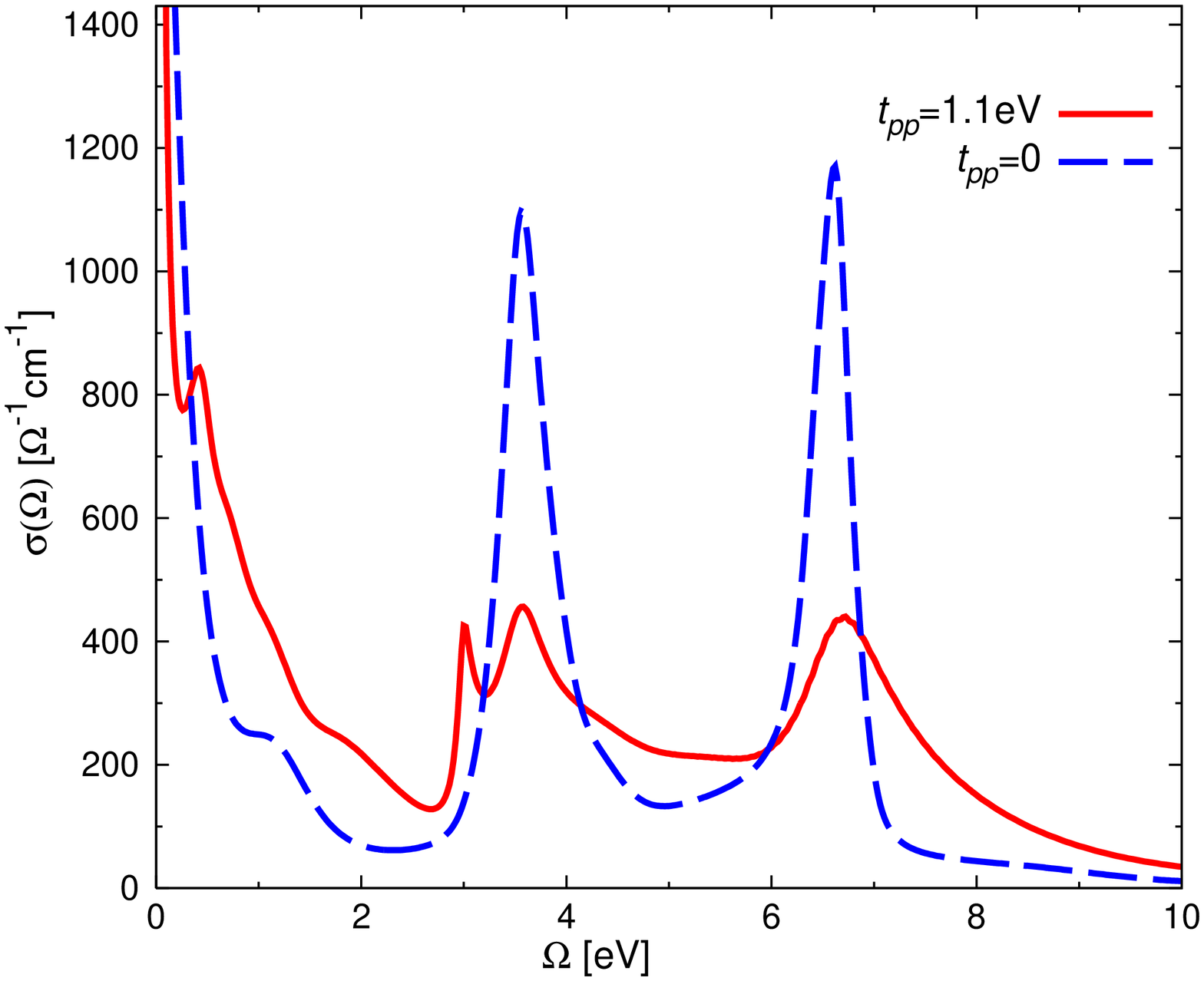}
(b)\includegraphics[width=6.5cm, angle=-90]{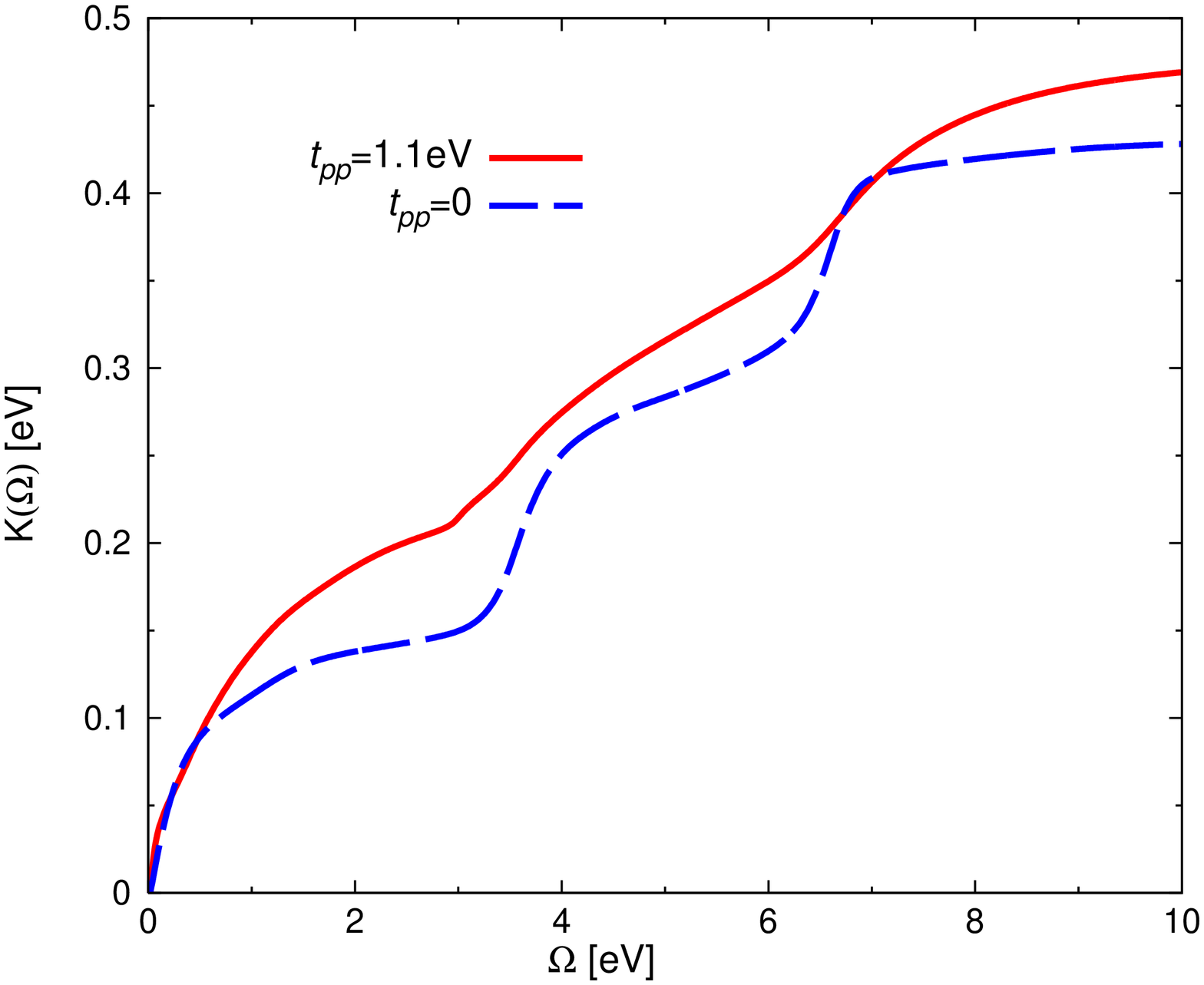}
    \caption{Optical conductivity [Panel (a)] and the corresponding integrated spectral weight [Panel (b)]
    in the 0.18 hole doped PM case. Parameters for $t_{pp}=1.1$ eV: $\Delta=4.2$ eV, $\varepsilon_d=-7.29$ eV,
    $\varepsilon_p=-3.09$ eV;
    Parameters for $t_{pp}=0$: $\Delta=4.5$ eV, $\varepsilon_d=-7.93$ eV, $\varepsilon_p=-3.43$ eV. Common parameters: $U=9$ eV, $t_{pd}=1.6$ eV, $T=0.1$ eV.}
    \label{condPM}
\end{figure}

Fig.~\ref{condPM} shows the optical conductivities and integrated spectral weights for 0.18 hole-doping case.  For $t_{pp}=0$ we had previously found \cite{demedici09} that hole-doping (but not electron-doping) activated a sharp, large amplitude  peak at $3.5$ eV associated with transitions from the non-bonding oxygen band to the Zhang-Rice holes created by doping. We see that inclusion of a non-vanishing oxygen-oxygen hopping reduces the amplitude of this peak (which is not observed experimentally), spreading the weight over a range of frequencies. Interestingly we see that the oscillator strength in the  ``Drude'' zero-frequency peak is essentially independent of $t_{pp}$, while differences appear at frequencies larger than about $0.7$ eV. This suggests that while a comparison of the low-frequency optical conductivity data to model systems may be meaningful, the conductivity at higher frequencies (but still below the charge transfer gap) depends strongly on details of the model.  Note that in the single-site approximation employed here the low-frequency spectral weight is proportional to the quasiparticle renormalization $Z$. The negligible $t_{pp}$-dependence seen in the low-frequency part of the conductivity integral therefore confirms the negligible $t_{pp}$ dependence of $Z$ in this doping range. 

\subsection{Comparison to one-band model}

\begin{figure}
    \centering
(a)\includegraphics[width=6.5cm, angle=-90]{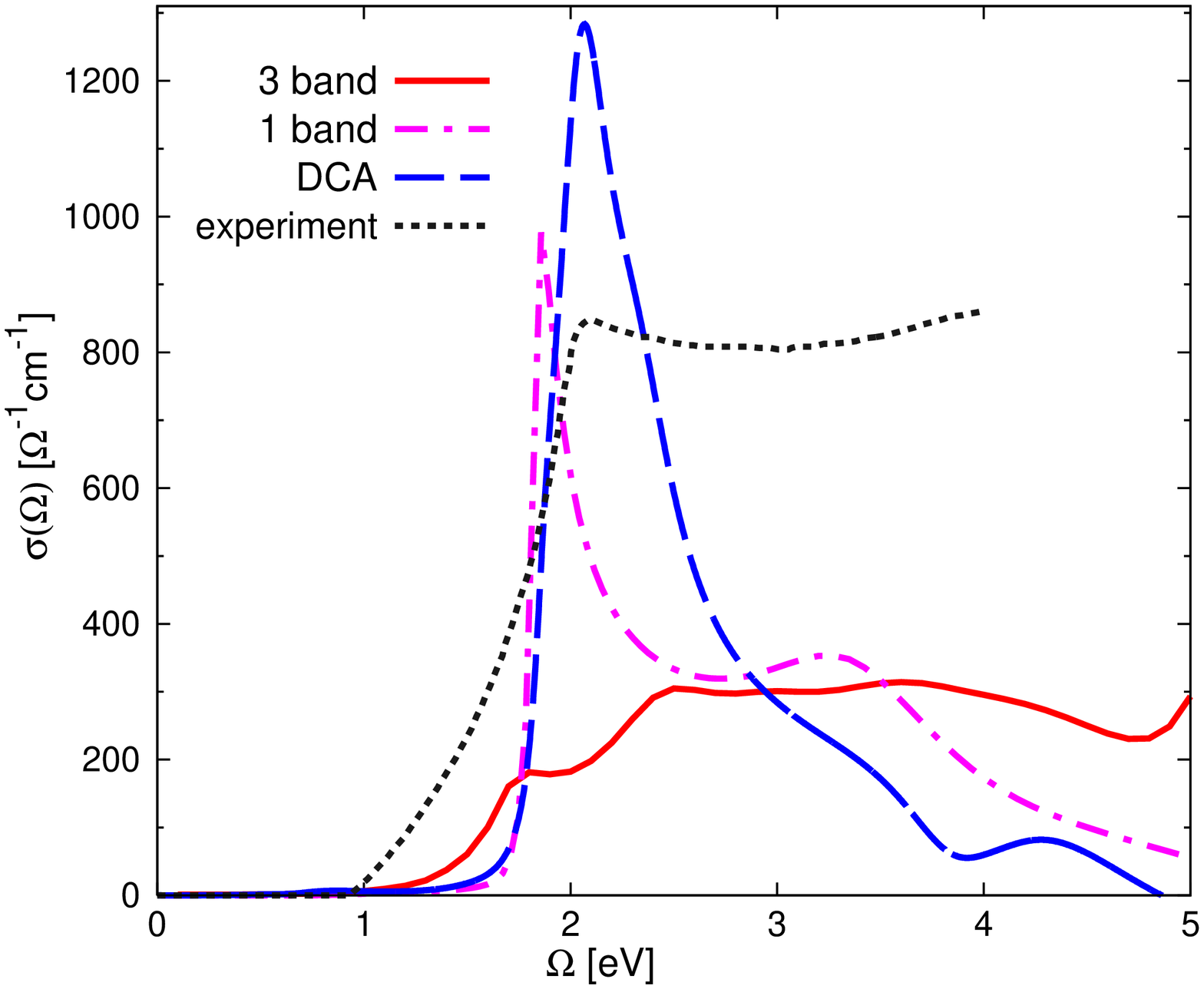}
(b)\includegraphics[width=6.5cm, angle=-90]{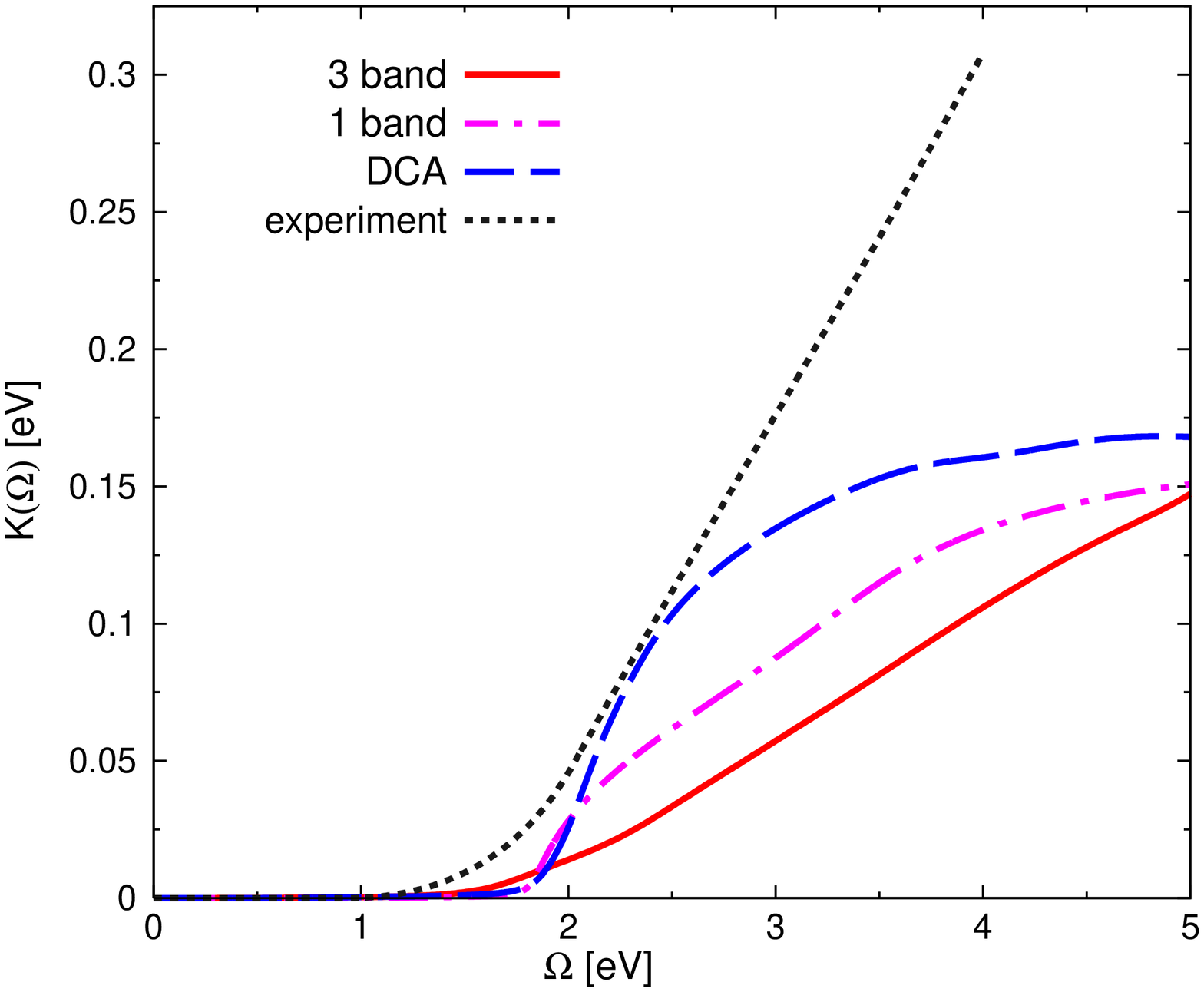}
    \caption{
    Comparison of optical conductivities [Panel (a)] and integrated spectral weights [Panel (b)] between the three-band calculation, one-band calculations and experiments in the undoped case. Solid lines (red online): $t_{pp}=1.1$ eV three-band calculation, parameters as the upper panel of Fig.~\ref{dosAFM}. Dash-dotted lines (magenta online): single-site DMFT calculation of the one-band model [Eq.~\eqref{H1band}] in the AFM phase. Dashed lines (blue online): four-site DCA calculation\cite{Lin09}
    of the one-band model [Eq.~\eqref{H1band}] in PM phase. Parameters for one-band calculations: $t=0.37$ eV, $t'=-0.3t$, $U_{\rm eff}=9t$, $\varepsilon_d=-4.5t$, $T=0.1t$. Dotted lines: experiments of Ref.~\onlinecite{Uchida91}.}
    \label{comp1band}
\end{figure}

Fig.~\ref{comp1band} compares the $t_{pp}=1.1$ eV conductivity calculation of the undoped AFM case to two calculations of the conductivity of the  one-band Hubbard model [Eq.~\eqref{H1band}]. The ``1 band'' curve (dash-dotted line, magenta online) denotes a single-site DMFT calculation of the one-band model in AFM phase, with $U$ adjusted to give the same gap. One sees that the conductivity at the gap edge is much more sharply peaked in the one-band than in the three-band model. The ``DCA'' curve (dashed line, blue online) denotes a four-site-cluster DCA calculation of the one-band model in PM phase.\cite{Lin09} This calculation includes short range AFM correlations, which are seen to act further to steepen the line-shape and pile up even more weight near the gap edge. Performing a cluster calculation for the three-band model would be very valuable. Despite all these obvious differences, we see that the integrated spectral weight of the three cases are similar at around 5 eV, above which the oxygen band below $\varepsilon_p$ participates in the optical transition which is not contained in any one-band calculations. In other words, an effective one-band model describes the degrees of freedom relevant to the conductivity up to the scale of $\sim 5$ eV but even in this frequency range the matrix elements and detailed structure of the conductivity require information which is beyond the scope of the one-band model.

\begin{figure}
    \centering
(a)\includegraphics[width=6.5cm, angle=-90]{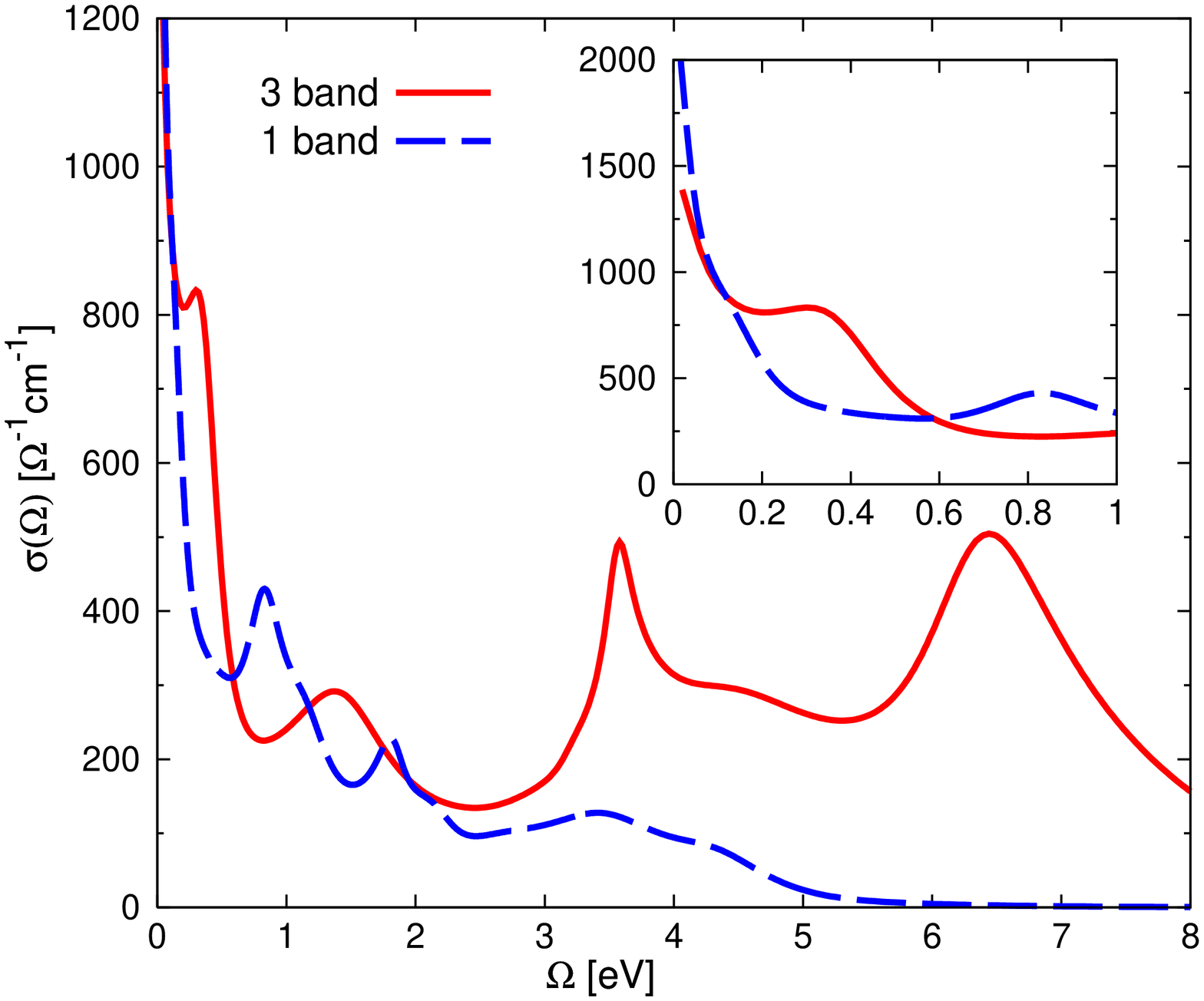}
(b)\includegraphics[width=6.5cm, angle=-90]{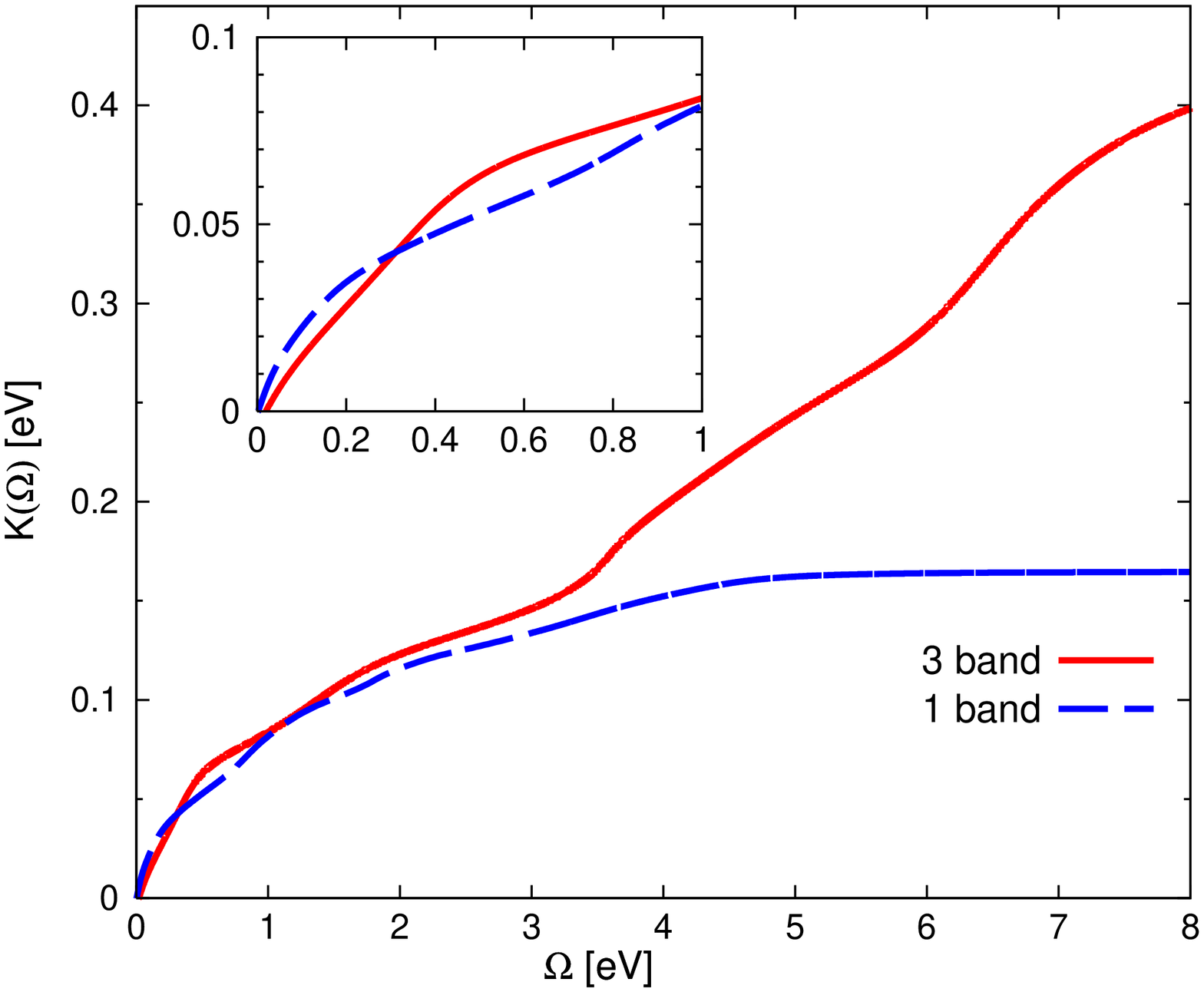}
    \caption{
    Comparison of optical conductivities [Panel (a)] and integrated spectral weights [Panel (b)] between the three-band calculation and one-band calculation in the 0.10 hole-doped case. Solid lines (red online): three-band calculation. Dashed lines (blue online): single-site DMFT calculation of the one-band model [Eq.~\eqref{H1band}]. Both calculations are in the PM phase. Insets: zoom-in of the low-frequency part.
Parameters for three-band calculations: $\Delta=4.2$ eV, $\varepsilon_d=-7.64$ eV, $\varepsilon_p=-3.44$ eV, $U=9$ eV, $t_{pd}=1.6$ eV, $t_{pp}=1.1$ eV, $T=0.1$ eV.
Parameters for one-band calculations: $t=0.37$ eV, $t'=-0.3t$, $U_{\rm eff}=9t$, $\varepsilon_d=-2.42t$, $T=0.1t$.}
    \label{comp1band-0p1}
\end{figure}

Fig.~\ref{comp1band-0p1} compares the optical conductivity calculated from three-band model and one-band model at 0.10 hole doping. At this doping value both calculations are in PM phase. We see that the one-band model provides a reasonable description up to around 3 eV, above which transition from the oxygen band to Zhang-Rice band appears and the one-band model result deviates from the three-band model one. Examination of the insets (which are the zoom-in of the low-frequency regime) shows that although the detailed line shape of the Drude peaks are slightly different, their Drude weights are quite similar.

\section{ Comparison to Ref.~17}\label{comparison}

\begin{figure}[]
    \centering
(a)\includegraphics[width=6cm, angle=-90]{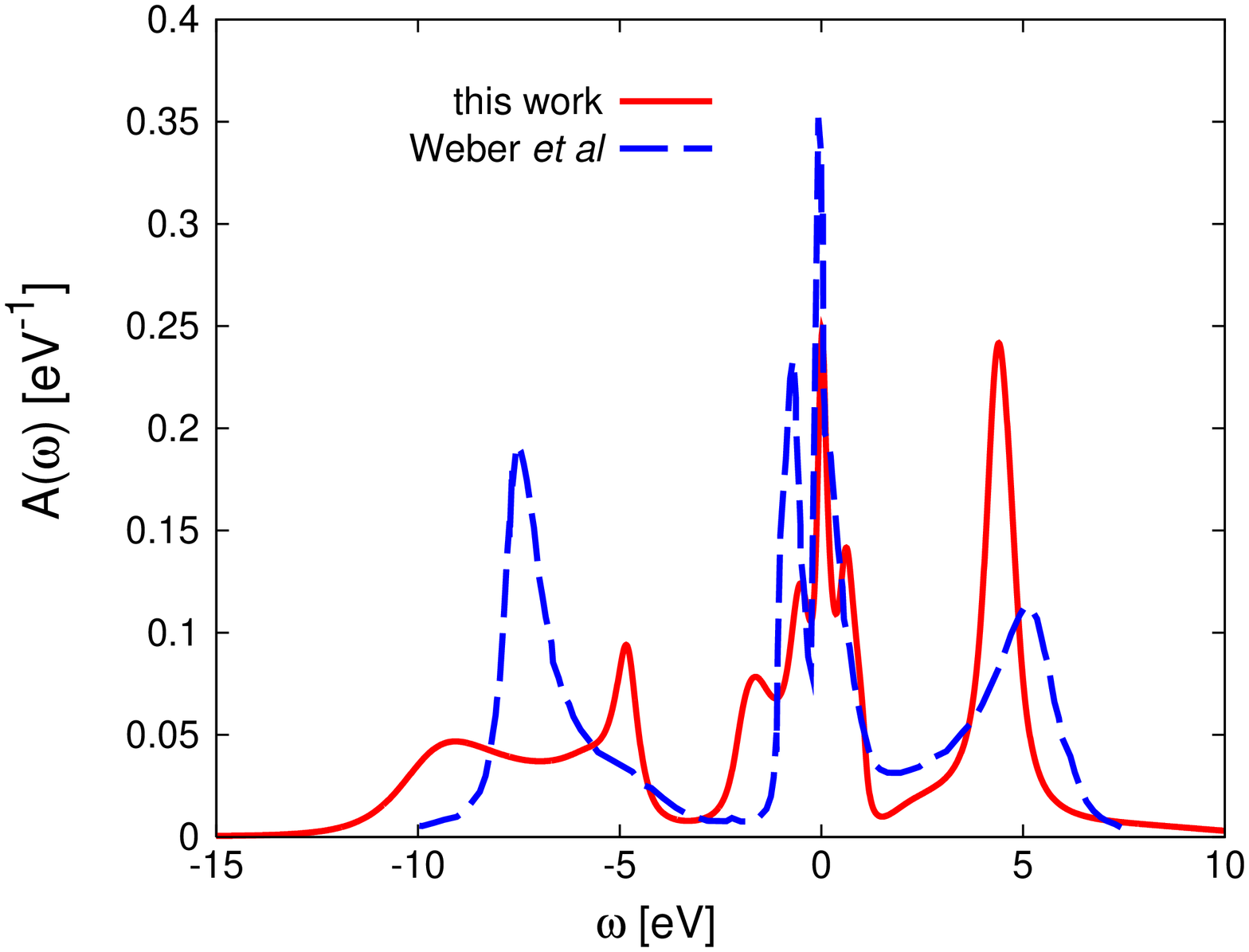}
(b)\includegraphics[width=6cm, angle=-90]{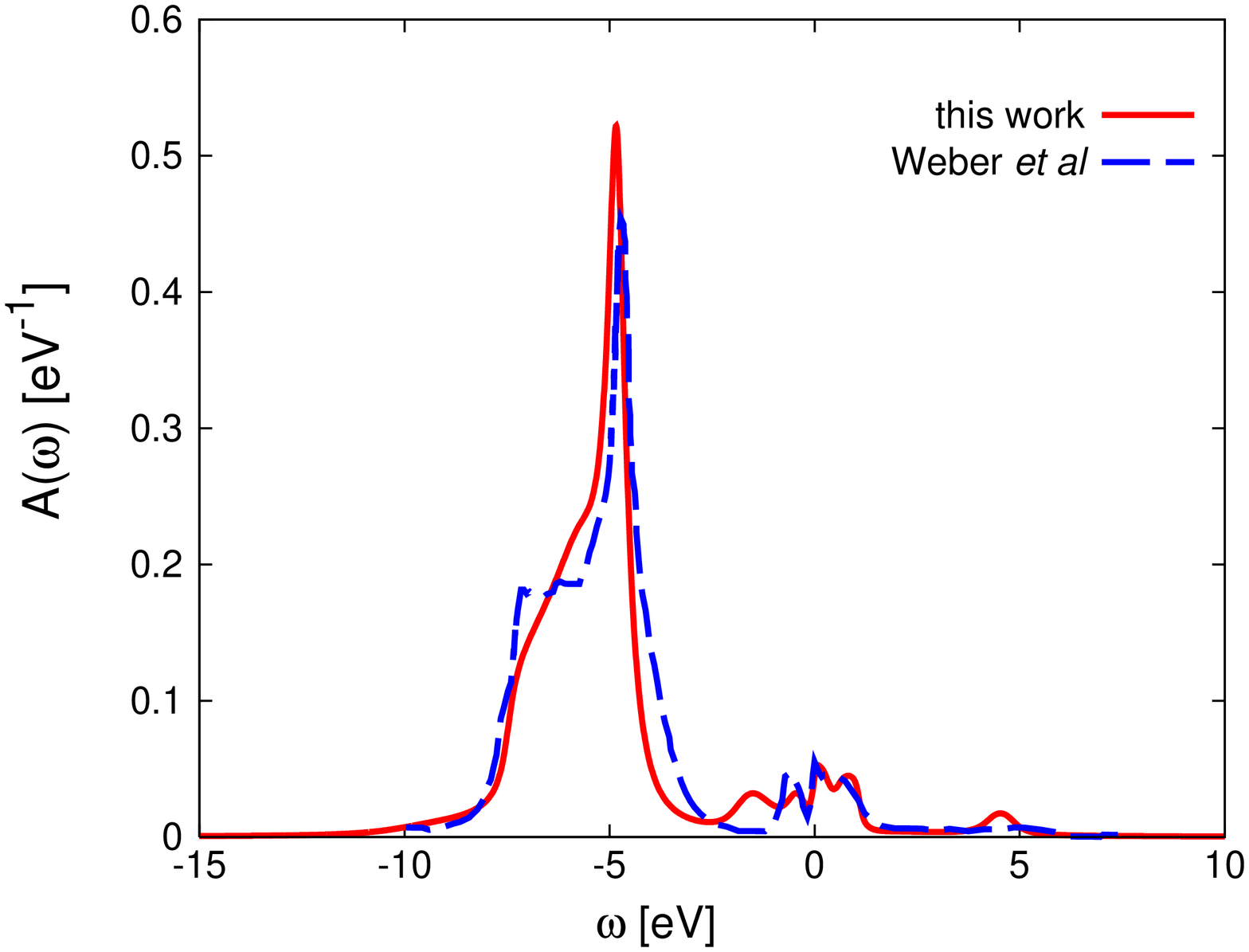}
    \caption{Comparison of 0.24 hole doping density of states obtained for the model of Ref.~\onlinecite{Weber08} in this work (solid line, red on-line) and digitized from Ref.~\onlinecite{Weber08} (dashed line, blue on-line). (a) Copper density of states. (b) Oxygen density of states. Parameters: $t_{pd}=1.41$ eV, $t_{pp}=0.66$ eV, $U=8$ eV, $\Delta=0.34$ eV, $\varepsilon_d=-5.17$ eV, $\varepsilon_p=-4.83$ eV, $T=0.1$ eV (this work) and $T \approx 0.008$ eV (Ref.~\onlinecite{Weber08}) .}
    \label{compWeberDoS}
\end{figure}

\begin{figure}[]
    \centering
    \includegraphics[width=6cm, angle=-90]{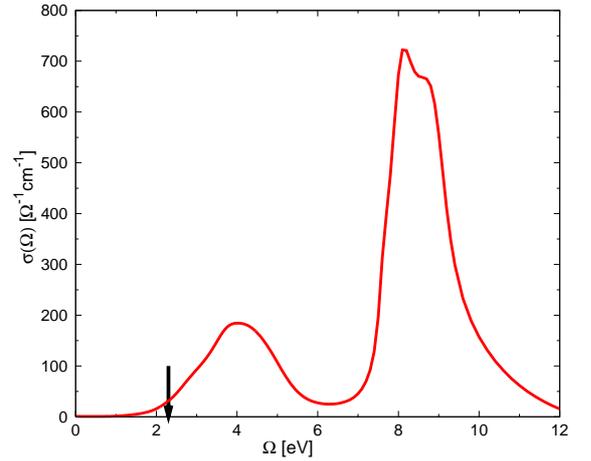}
    \caption{Optical conductivity for undoped paramagnetic insulator calculated for the model of Ref.~\onlinecite{Weber08}. Parameters: $\varepsilon_d=-7.5$ eV, $\varepsilon_p=-7.16$ eV at temperature  $T=0.1$ eV. The arrow indicates the gap value of approximately 2.3 eV calculated from the quasiparticle equation of Ref.~\onlinecite{Wang09}.
 }
    \label{compWebercondPMI}
\end{figure}

\begin{figure}[]
    \centering
    \includegraphics[width=6cm, angle=-90]{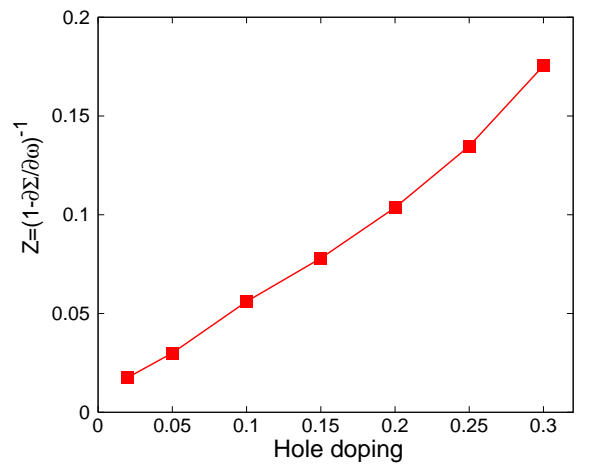}
    \caption{Quasiparticle weight $Z=(1-\partial\Sigma/\partial\omega)^{-1}|_{\omega\rightarrow0}$ as a function of doping calculated using an ED solver at zero temperature, for the model of Ref.~\onlinecite{Weber08}.}
    \label{Zplot}
\end{figure}

\begin{figure}[]
    \centering
(a)\includegraphics[width=6cm, angle=-90]{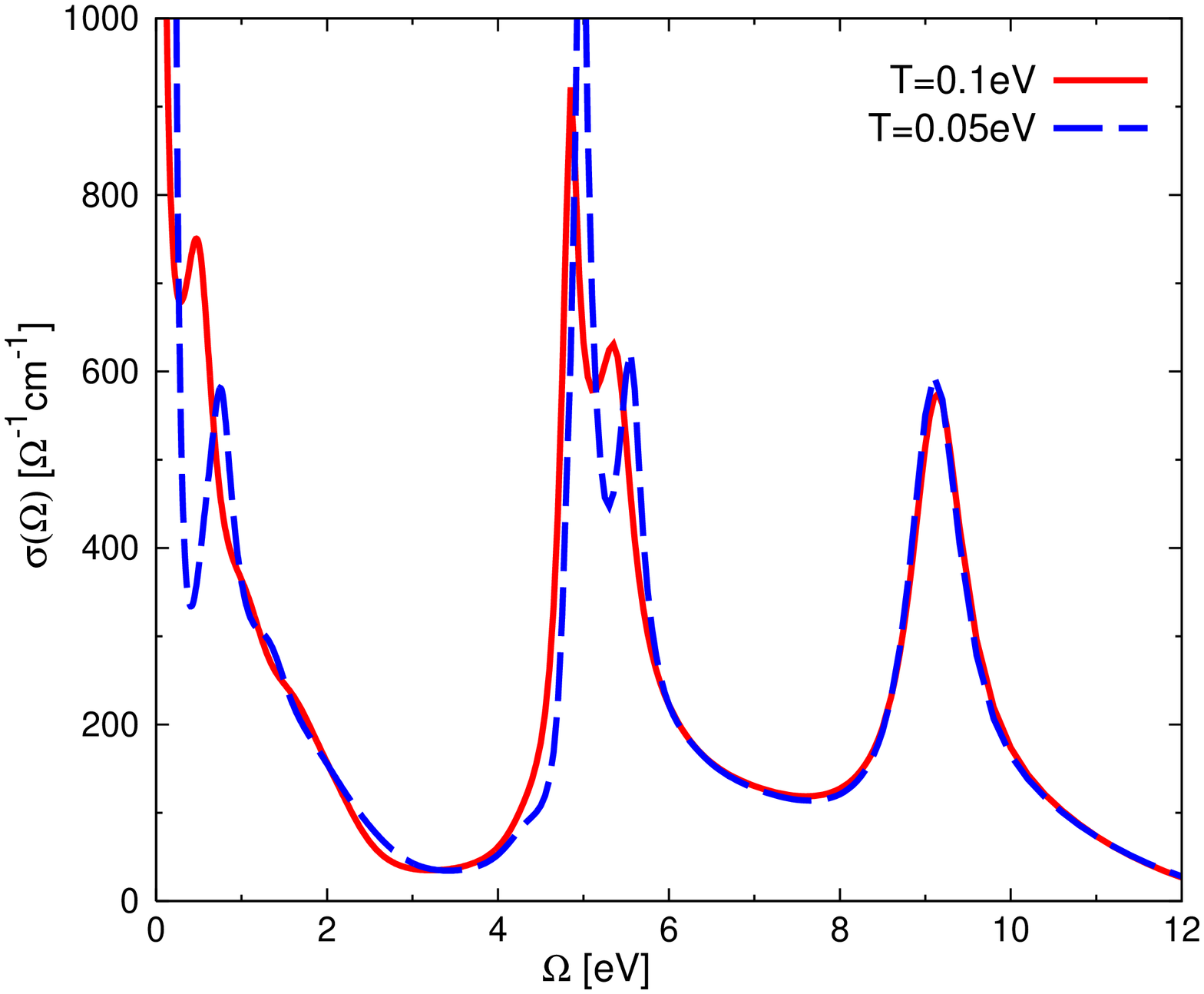}
(b)\includegraphics[width=6cm, angle=-90]{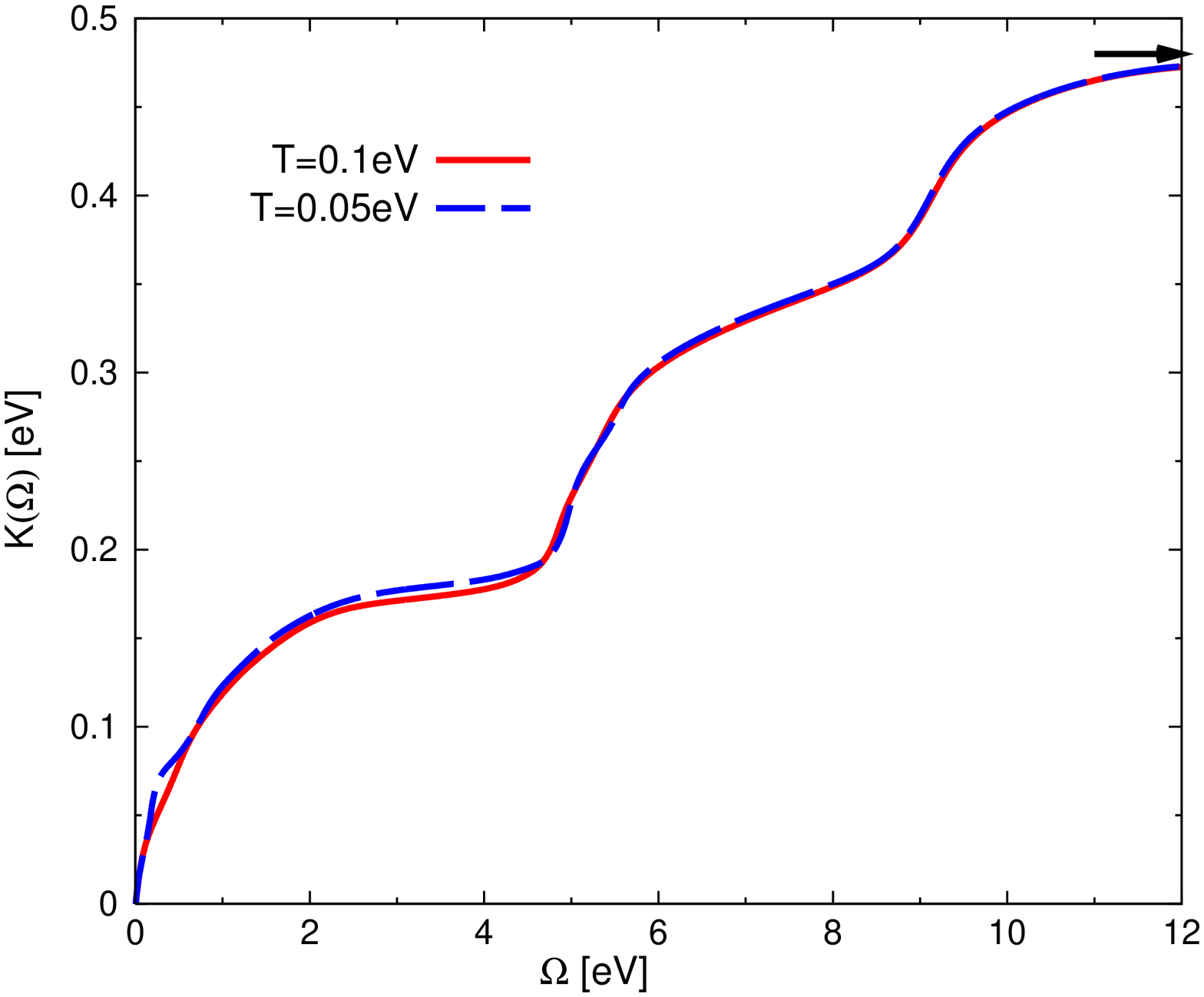}
    \caption{Optical conductivities [panel (a)] and integrated spectral weight [panel (b)] at 0.24 hole doping of two different temperatures calculated for the model of Ref.~\onlinecite{Weber08}. Parameters: $\varepsilon_d=-5.17$ eV at $T=0.1$ eV; $\varepsilon_d=-5.22$ eV at $T=0.05$ eV. The arrow at the upper right corner of panel (b) indicates the kinetic energy $K=0.48$ eV independently calculated using Eq.~\eqref{sumrule}.}
    \label{compWebercond0p24}
\end{figure}

In this section we turn to a detailed examination of the specific parameter values reported in Ref.~\onlinecite{Weber08} to describe La$_{2-x}$Sr$_x$CuO$_4$. These parameters place La$_2$CuO$_4$ in the strongly correlated, paramagnetic insulator regime of the phase diagram. We are interested in the gap value in the insulating state, the conductivity in the doped metallic state and the doping dependence of the electron quasiparticle weight. By comparing these results to experiment we can assess whether the parameters proposed in Ref.~\onlinecite{Weber08} in fact provide a reasonable starting point for a description of  La$_{2-x}$Sr$_x$CuO$_4$.  As a by-product we will see that whether one uses the form of the oxygen-oxygen Hamiltonian given in our Eq.~\eqref{Hpm}  or the form given  in Refs.~\onlinecite{Hybertsen89,Mila88b,Korshunov05,Weber08,Weber10a,Weber10b,Hanke10} does not affect the physics in any significant way.

We first verify that the model studied here is the same as that studied in Ref.~\onlinecite{Weber08}. Verification is necessary because Ref.~\onlinecite{Weber08}  used a tight-binding model which was obtained from a detailed fit to a band theory calculation and differs in two ways from our Eq.~\eqref{Hpm}. First, even longer-ranged oxygen-oxygen hoppings are included. However, the amplitudes of these long ranged terms are very small (less than 5\% of the basic $t_{pp}$) and (as we shall see below in Fig.~\ref{compWeberDoS}) have a negligible effect on the oxygen density of states. Second, the  diagonal elements of the hopping  do not contain the $t_{pp}$ term so our Eq.~\eqref{Hpm} must be changed so that the diagonal terms become ($\varepsilon_d$, $\varepsilon_p$, $\varepsilon_p$) and the current operator must be correspondingly modified. Ref.~\onlinecite{Weber08} chose $U=8$ eV.  The precise parameter values for $t_{pd}$ and $t_{pp}$ were not given but subsequent work\cite{Weber10b} indicates $t_{pd}=1.41$ eV and $t_{pp}=0.66$ eV. In Ref.~\onlinecite{Weber08} the values of $\varepsilon_d$ and $\varepsilon_p$ were obtained by applying a double-counting correction to results obtained from a band calculation and correspond in our notations to $\Delta=\varepsilon_p-\varepsilon_d=0.34$ eV.  We have performed calculations with the modified Eq.~\eqref{Hpm}, using these parameters, at temperatures $T=0.1$ eV and 0.05 eV.  Comparison to Fig.~\ref{phase} here shows that even for $t_{pd}=1.6$, these parameters would lie on the insulating side of the phase boundary; the smaller $t_{pd}$ would only make the model more insulating.

We have performed several tests to verify the consistency of the models studied here and in Ref.~\onlinecite{Weber08}. First we have computed the $d$-level occupancy and total density as functions of chemical potential $n_d(\mu)$ and $n_{\rm tot}(\mu)$ and have verified that they agree with those presented by Ref.~\onlinecite{Weber08} to within 2\%.  The projection of our calculated many-body density of states onto the oxygen orbitals is shown in panel (b) of Fig.~\ref{compWeberDoS}, and is seen to correspond closely to the results obtained by digitizing the figures in  Ref.~\onlinecite{Weber08}. The Cu-projected densities of states [shown in panel (a) of Fig.~\ref{compWeberDoS}] are qualitatively similar. The difference in peak height near $\omega=0$ is probably a reflection of the different temperatures studied. The difference in widths of the higher energy features may reflect unavoidable uncertainties in analytic continuation (which is insensitive to fine details of high energy features). We note however that the integral from $-15$ eV to $0$ of our $A_d$ curve reproduces the directly measured $d$ occupancy $n_d\approx 0.53$  while the integral of the digitized curve from Ref.~\onlinecite{Weber08} yields $n_d \approx 0.58$. We therefore suspect that our continuation is more accurate.

We have not been able to stabilize the AFM insulating state at the temperatures accessible to us, so we confine our discussion to PM states. Our calculated conductivity for the undoped paramagnetic insulator is shown in Fig.~\ref{compWebercondPMI}. We believe that the broadening of the gap edge is in part an artifact of analytic continuation, and in part a temperature effect. As shown in Ref.~\onlinecite{Wang09} the gap may be estimated from the difference between the highest negative frequency and lowest positive frequency at which the quasiparticle equation $\omega-{\rm Re}\Sigma(\omega)-\varepsilon_k=0$ is satisfied. This analysis yields a gap of 2.3 eV shown as the arrow in Fig.~\ref{compWebercondPMI}. This gap is larger than the 1.8 eV found experimentally and it is difficult to see how adding antiferromagnetism to the present calculation will do anything but increase the gap. This analysis therefore suggests  that the parameters employed in Ref.~\onlinecite{Weber08} overestimate the correlation strength in La$_2$CuO$_4$. However, it must be emphasized that the extraction of gap values from DMFT data in insulating states is complicated by the intrinsic difficulties of analytic continuation (for QMC data) and discretization issues (for ED data) and that these difficulties become particularly severe for strong correlations (implying large gaps). While we believe that the quasiparticle equation method of Ref.~\onlinecite{Wang09} is the most reliable method for extracting gaps, the issue may not yet be settled.

We next turn to the hole-doped metallic state. Fig.~\ref{Zplot} shows the doping dependence of the quasiparticle weight $Z=(1-\partial\Sigma/\partial\omega)^{-1}$ computed using the $T=0$ ED method. The small value and approximately linear doping dependence are hallmarks of a strongly correlated metal. The temperature dependent ED method (result not shown) indicates a weak increase of $Z$ as $T$ is increased, while analytic continuation of QMC data at 0.24 hole doping and $T=0.1$eV yields $Z\simeq0.16$. Fig.~3(b) of Ref.~\onlinecite{Weber08} presents results for $Z$ which are inconsistent with the results presented here, being larger and less strongly doping dependent ($Z\simeq0.3$ at hole doping $x=0.3$, decreasing to $Z\simeq0.22$ at $x=0.08$). The consistency we have found, between QMC and ED, leads us to believe that the results presented here are correct.  

These results are very similar to those shown at the $\Delta=2$ eV hole doping curve (black squares) in Fig.~3 of Ref.~\onlinecite{demedici09} for the model with $t_{pp}=0$, indicating that the doping dependence is the same ($Z\sim x$), but the values reported for the $t_{pp}=0$ case are larger ($Z\sim 0.75x$) as expected because the $t_{pp}=0$ calculation was performed for slightly weaker correlations. This indicates that the oxygen-oxygen hopping does not have an important effect on this aspect of the physics.

The $Z$ values presented here imply a (zone-diagonal) Fermi velocity $v_F\simeq1{\textrm{eV-\AA}}$ at $x=0.24$, decreasing approximately linearly with doping $x$. The experimental values are around 2eV, independent of $x$ in the range $0.1<x<0.25$ and are consistent with values inferred, using Fermi liquid theory,  from specific heat and quantum oscillation measurements at dopings $x>0.2$ \cite{Vignolle08}. (Very recent experiments \cite{Plumb10} find a further $30\%$ decrease in $v_F$ at low $T$ and $\omega<10$ meV; this is still strongly inconsistent with our calculations). We believe the difference between calculation and data arises because the correlations in the actual material are less strong than assumed in Ref.~\onlinecite{Weber08}.
 
Panel (a) of Fig.~\ref{compWebercond0p24} shows the conductivity calculated using analytically continued QMC results at hole doping $x=0.24$ at two temperatures. Three checks can be made to our result. First, we have computed $\int_0^\infty2d\Omega\sigma(\Omega)/\pi$ both directly from $\sigma(\Omega)$ [panel (b) of Fig.~\ref{compWebercond0p24}] and from Eq.~\eqref{sumrule} [finding $K=0.48$ eV, shown as an arrow in panel (b) of Fig.~\ref{compWebercond0p24}].  The two results are consistent. Second, a straightforward unit conversion indicates that our $K(\omega=1.8{\rm eV})\simeq0.15{\rm eV}$ corresponds to $N_{\rm eff}\simeq 0.28$, in reasonable agreement with the estimate $N_{\rm eff}\simeq0.29$ inferred from Fig.~3(a) of  Ref.~\onlinecite{Weber08}. Third, the low-frequency conductivity is characterized by a ``Drude'' peak, whose area may be obtained (if vertex corrections are negligible) from the average of the Fermi velocity over the Fermi surface:
\begin{equation}
K_{\rm Drude}=\int\frac{ds}{4\pi^2}\left|v_F^*(s)\right|\label{Kdrudedeftext}
\end{equation}
with $s$ a coordinate along the Fermi surface and $v_F^*(s)$ the renormalized Fermi velocity (see Appendix for details). We evaluated $K_{\rm Drude}$ from the product of the height and half-width of the zero frequency peak, obtaining 0.06 eV and from Eq.~\eqref{Kdrudedeftext} obtaining 0.07 eV. We have also repeated the conductivity calculation at the lower temperature $T=0.05$ eV; while the form of the conductivity in the mid-IR range somewhat changes, both the area in the Drude peak and the area at $\Omega<1$ eV remain essentially unchanged. 

Although as mentioned above the spectral weight integrated up to 1.8 eV falls on the curve plotted in Fig.~3(a) of Ref.~\onlinecite{Weber08}, our conductivity differs from that shown in Fig.~4 of Ref.~\onlinecite{Weber08}, being for example larger by a factor of around 2 in the 1-3 eV range. The authors of Ref.~\onlinecite{Weber08} inform us\cite{Weber.comm} that there may be a normalization error in the result.

We now compare the results calculated here to those obtained in experiment. From panel (b) of Fig.~\ref{compWebercond0p24} we see that $K(\Omega=0.8 {\rm eV})\simeq0.1{\rm eV}\approx1.6K_{\rm Drude}$. We may obtain experimental estimates from Fig.~3 of Ref.~\onlinecite{Comanac08} and from Fig.~4.4 of Ref.~\onlinecite{ComanacPhD} (note that $K_{\rm band}$ in this reference $\approx0.4$ eV).  We find $K^{\rm exp}(\Omega=0.8 {\rm eV})\approx0.12\pm0.02{\rm eV}$ and $K^{\rm exp}_{\rm Drude}\approx0.06\pm0.01{\rm eV}$. We believe that the difference between the calculated and measured velocities means that the agreement of the conductivity is accidental.

To conclude this subsection we note that the results shown here calculated using the Hamiltonian with a slightly different form\cite{Weber08,Hybertsen89,Korshunov05,Hanke10,Mila88b} than previous sections can be compared to the results obtained using the model of Ref.~\onlinecite{Andersen95} in the rest of this paper. Comparison of Figs.~\ref{compWeberDoS} and \ref{compWebercond0p24} with Figs.~\ref{dosPM1} and \ref{condPM} reveals that the different choice of models does not change the qualitative nature of the spectrum and the optical conductivity, and the difference can be understood from effectively different values of $t_{pp}$.

\section{Conclusion}\label{conclusion}

In this paper we have considered the effect of oxygen-oxygen hopping $t_{pp}$ on the particle-hole asymmetry and optical conductivity of a three-band copper-oxide model related to cuprate superconductors. We have found that the inclusion of oxygen-oxygen hopping does not change the physics in any important way.  In particular, in the model with large oxygen-oxygen hopping, as in the model with no oxygen-oxygen hopping, parameters which place the model on the paramagnetic insulator side of the metal-insulator phase diagram lead to results for the insulating gap and, most importantly, the quasiparticle mass, which are in disagreement with experiment, suggesting that La$_2$CuO$_4$ is not as strongly correlated as may have been believed. A model with weaker correlations would yield a larger and less strongly doping dependent Fermi velocity while vertex corrections (neglected in the single-site DMFT but shown via phenomenological analysis to be important in the cuprates \cite{Millis03, Millis05}) could produce the correct behaviour of the conductivity.

Incorporation of oxygen-oxygen hopping does not resolve the most dramatic discrepancy between single-site DMFT theory of the  three-band model   and experiment, namely the experimentally observed conductivity near the insulating gap edge is much larger than that predicted theoretically. Resolving the disagreement probably requires the introduction of additional bands, as noted in Ref.~\onlinecite{Weber10b}. However, inclusion of oxygen-oxygen hopping helps resolve a different problem, by broadening a peak at $\sim3.5$ eV  which is theoretically predicted to appear upon hole doping. This peak is not observed experimentally.

Inclusion of oxygen-oxygen hopping does have several important effects. As previously observed by  Ref.~\onlinecite{Weber08}, the use of a more realistic band structure brings the magnetic phase diagram into better agreement with experiment. Further, the broadening of the non-bonding oxygen band due to oxygen-oxygen hopping may cause the lowest-lying $d^8$ band to be absorbed in the oxygen band rather than being visible as a separate feature.

We have compared our three-band calculations to one-band calculations. In the undoped AFM case, the one-band model clearly overestimates the conductivity in the vicinity of the gap edge; however the optical spectral weights in the doped cases at frequencies below $\sim 0.8$ eV do agree. This suggests that a reduction of the three-band model to a one-band model in AFM case is only possible at energies well below the charge transfer gap, in agreement with previous remarks \cite{Sawatzky06,Weber10a} and in disagreement with our previous work.\cite{demedici09} In contrast, for PM case with doping, the one-band model does provide reasonable description up to around 3 eV.

We have also compared our results to those presented in Ref.~\onlinecite{Weber08}. We find, in disagreement with statements in this work, that the main origin of the differences between that work and ours is not related to the choice of oxygen-oxygen hopping Hamiltonian, but instead relates to calculational issues.  Our results suggest that the key issue in fitting a DMFT calculation to band theory is the value of   $\Delta\equiv\varepsilon_p-\varepsilon_d$. The value of $\Delta$ is affected by  the value chosen for the double counting correction, which is not on a firm theoretical basis.  

\section*{Acknowledgments}
We thank M. Capone, H. T. Dang and E. Gull for helpful discussions, and N. Lin for kindly providing the DCA data and for discussions and are particularly grateful to C. Weber for very extensive and helpful discussions concerning the differences between our results.   XW and AJM are financially supported by NSF-DMR-1006282 and LdM by Program ANR-09-RPDOC-019-01 and by RTRA Triangle de la Physique. Part of this research was conducted at the Center for Nanophase Materials Sciences, which is sponsored at Oak Ridge National Laboratory by the Division of Scientific User Facilities, U.S. Department of Energy.

\appendix
\renewcommand{\theequation}{A-\arabic{equation}}
\setcounter{equation}{0}
\section*{Appendix}

In this appendix we present an analysis of the low-frequency conductivity, based on the assumptions that the self-energy is momentum independent and the model is in a Fermi liquid regime. We shall show that on these assumptions the Drude weight is given by an integral of the Fermi velocity over the Fermi surface.  To do this we first note that in the low-frequency limit of the Fermi liquid regime the self-energy (in general, a matrix in the space of band indices which we denote in bold-faced type) can be written
\begin{equation}
{\bf \Sigma}(\omega)=-{\mathbf \Delta_\mu}+i{\bf \Gamma} +\left({\bf 1}-{\bf Z}^{-1}\right)\omega
\label{Sigmalow}
\end{equation}
Here we have assumed that there is no momentum dependence.  ${\mathbf \Delta_\mu}$ in principle both re-arranges the bands and acts as an effective shift of chemical potential; it is  given  by the zero frequency limit of the  real part of the self-energy.  ${\mathbf \Gamma}$ is the level broadening given by the imaginary part of the self-energy. In Fermi liquid regime ${\mathbf \Gamma}$ is small. ${\mathbf Z}$ is a mass renormalization factor which relates to the slope of the real part of the self-energy at the Fermi energy.

We may rewrite Eq.~\eqref{Gdef} as
\begin{equation}
{\bf G}^{-1}(\omega, \boldsymbol{k})={\bf Z}^{-1/2}\left[\omega-{\bf H}^*(\boldsymbol{k})-i{\mathbf \Gamma}^*\right]{\bf Z}^{-1/2}
\label{GFL}
\end{equation}
with 
\begin{eqnarray}
{\bf H}^*(\boldsymbol{k})&=&{\bf Z}^{1/2}\left({\bf H}(\boldsymbol{k})-{\mathbf \Delta_\mu}\right){\bf Z}^{1/2}
\label{Hstardef} \\
{\bf \Gamma}^*&=&{\bf Z}^{1/2}{\mathbf \Gamma}{\bf Z}^{1/2}
\label{Gammastardef}
\end{eqnarray}
Thus in analogy to Eq.~\eqref{Adef} we define ${\mathbf A}^*$ as 
\begin{align}
{\mathbf A}^*&=\frac{\left[\omega-{\bf H}^*(\boldsymbol{k})-i{\mathbf \Gamma}^*\right]^{-1}-\left[\omega+{\bf H}^*(\boldsymbol{k})+i{\mathbf \Gamma}^*\right]^{-1}}{2i}\nonumber\\
&={\bf Z}^{-1/2}{\bf A}{\bf Z}^{-1/2}
\label{Astardef}
\end{align}
and the renormalized current operator ${\bf j}^*$ as
\begin{equation}
{\bf j}^*(\boldsymbol{k})={\bf Z}^{1/2}\frac{\partial {\bf H}(\boldsymbol{k})}{\partial \boldsymbol{k}}{\bf Z}^{1/2}\equiv \frac{\partial {\bf H}^*(\boldsymbol{k})}{\partial \boldsymbol{k}}
\label{jstardef}
\end{equation}

The the expression of the optical conductivity [Eq.~\eqref{sigmamatrix}] can be straightforwardly rewritten as
\begin{align}
\sigma(\Omega)&=\frac{2e^2}{\hbar c}\int_{-\infty}^\infty \frac{d\omega}{\pi}\int
\frac{d^2\boldsymbol{k}}{(2\pi)^2}\frac{f(\omega)-f(\omega+\Omega)}{\Omega} \nonumber
\\
&\times{\mathrm{Tr}}\left[{\bf j}^*(\boldsymbol{k}){\bf A}^*(\omega+\Omega,\boldsymbol{k}){\bf
j}^*(\boldsymbol{k}){\bf A}^*(\omega,\boldsymbol{k})\right], \label{sigmastardef}
\end{align}

Eq.~\eqref{sigmastardef} holds if the self-energy is momentum independent and all frequencies are low enough that the Fermi liquid approximation is valid.

Because ${\mathbf Z}$ has only positive eigenvalues, the Fermi surface is the locus of $\boldsymbol{k}$ points $\boldsymbol{k}=\boldsymbol{k}_F$ for which ${\bf H}(\boldsymbol{k})-{\mathbf \Delta_\mu}$ has a zero eigenvalue. Labelling the eigenvalue of ${\bf H}$ which passes through zero at $\boldsymbol{k}=\boldsymbol{k}_F$   by $\varepsilon_{\boldsymbol{k}}$, we define the bare Fermi velocity $\boldsymbol{v}_F=\partial \varepsilon_{\boldsymbol{k}}/\partial \boldsymbol{k}$. To each point on the Fermi surface there corresponds a wavefunction $\psi^{\rm FS}_{\boldsymbol{k}}$  and the  Hellman-Feynman theorem implies that
\begin{equation}
\boldsymbol{v}_F\cdot {\bf \delta \boldsymbol{k}}=\left\langle\psi^{\rm FS}_{\boldsymbol{k}}\left|{\bf  H}(\boldsymbol{k})-{\bf H}(\boldsymbol{k}_F)\right|\psi^{\rm FS}_{\boldsymbol{k}}\right\rangle
\end{equation}
with $\delta\boldsymbol{k}=\boldsymbol{k}-\boldsymbol{k}_F$.

Similarly, at the Fermi surface one of the eigenvalues $E(\boldsymbol{k})$ of ${\bf H}^*(\boldsymbol{k})$ vanishes and corresponding to this eigenvalue is an eigenvector
\begin{equation}
\Psi^{\rm FS}_{\boldsymbol{k}}={\bf Z}^{-1/2}\psi^{\rm FS}_{\boldsymbol{k}}/\sqrt{Z^{-1}_{11}(\boldsymbol{k})}
\end{equation}
with normalization factor
\begin{equation}
Z^{-1}_{11}(\boldsymbol{k})=\left\langle\psi^{\rm FS}_{\boldsymbol{k}}|{\bf Z}^{-1}|\psi^{\rm FS}_{\boldsymbol{k}}\right\rangle.
\end{equation}
According to the Hellman-Feynman theorem, if we define the renormalized velocity $\boldsymbol{v}^*_F$ by 
\begin{equation}
\boldsymbol{v}^*_F\cdot\delta\boldsymbol{k}=\left\langle\Psi^{\rm FS}_{\boldsymbol{k}}\left|{\bf H}^*(\boldsymbol{k})-{\bf H}^*(\boldsymbol{k}_F)\right|\Psi^{\rm FS}_{\boldsymbol{k}}\right\rangle
\end{equation}
then 
\begin{equation}
\boldsymbol{v}^*_F(\boldsymbol{k})=\frac{\boldsymbol{v}_F(\boldsymbol{k})}{Z^{-1}_{11}(\boldsymbol{k})}
\label{vstar}
\end{equation}

Note that the renormalization of the velocity with respect to the band velocity is $\boldsymbol{k}$-dependent, even though the self-energy is not, essentially because the correlated orbital mixes differently with the uncorrelated orbitals depending on what Fermi surface point one considers.

Returning now to Eq.~\eqref{sigmastardef} we see that if ${\bf \Gamma}$ is sufficiently small then the dominant term in the conductivity is obtained by projecting everything onto the Fermi surface wave function so that (at small $\Omega$) 
\begin{align}
\sigma_{\rm qp}(\Omega)&=\frac{2e^2}{\hbar c}\int_{-\infty}^\infty \frac{d\omega}{\pi}\int
\frac{d^2\boldsymbol{k}}{(2\pi)^2}\frac{f(\omega)-f(\omega+\Omega)}{\Omega} \nonumber
\\
&\!\!\!\times j^*_{\rm FS}(\boldsymbol{k})A^*_{\rm FS}(\omega+\Omega,\boldsymbol{k})j^*_{\rm FS}(\boldsymbol{k})A^*_{\rm FS}(\omega,\boldsymbol{k}), \label{sigmaFSdef}
\end{align}
with $j^*_{\rm FS},A^*_{\rm FS}=\left\langle\Psi^{\rm FS}_{\boldsymbol{k}}\right|{\bf j}^*,{\bf A}^*\left|\Psi^{\rm FS}_{\boldsymbol{k}}\right\rangle$ the projections onto the Fermi surface of the current operator and spectral function. ``qp'' stands for ``quasi-particle''. For convenience we also define $\Gamma^*_{\rm FS}=\left\langle\Psi^{\rm FS}_{\boldsymbol{k}}\right|{\bf \Gamma}^*\left|\Psi^{\rm FS}_{\boldsymbol{k}}\right\rangle$

In the Fermi liquid limit we have
\begin{align}
j^*_{\rm FS}(\boldsymbol{k})&=v^*_{F,x}(\boldsymbol{k})\label{jFS}\\
A^*_{\rm FS}(\omega,\boldsymbol{k})&=\frac{\Gamma^*_{\rm FS}}{\left(\omega-\boldsymbol{v}^*_F\cdot\boldsymbol{\delta k}\right)^2+\left(\Gamma_{\rm FS}^*\right)^2}
\label{AFS}
\end{align}
Note that in Eq.~\eqref{sigmamatrix} we assumed that the current is in $x$-direction thus the $x$-component of the Fermi velocity is taken in Eq.~\eqref{jFS}. Plugging in Eq.~\eqref{sigmaFSdef} we have (at small $\Omega$) 
\begin{align}
\sigma_{\rm qp}(\Omega)&=\frac{2e^2}{\hbar c}\int
\frac{d^2\boldsymbol{k}}{(2\pi)^2}
\delta\left[\boldsymbol{v}^*_F\cdot(\boldsymbol{k}-\boldsymbol{k}_F)\right]\nonumber
\\
&\times\left[v^*_{F,x}(\boldsymbol{k})\right]^2\frac{2\Gamma^*_{\rm FS}}{\Omega^2+4\left(\Gamma^*_{\rm FS}\right)^2}
\end{align}
Thus just as in the usual case the integral of the low-frequency (Drude) part of the conductivity is given by the average over the Fermi surface of the renormalized Fermi velocity. Writing $d^2\boldsymbol{k}\rightarrow kdkd\theta$ we have (at small $\Omega$) 
\begin{align}
\sigma_{\rm qp}(\Omega)&=\frac{2e^2}{\hbar c}\int
\frac{d\theta k_F(\theta)}{4\pi^2}
\frac{v^*_{F,x}(\theta)^2}{\left|v^*_{F}(\theta)\right|}\frac{2\Gamma^*_{\rm FS}}{\Omega^2+4\left(\Gamma^*_{\rm FS}\right)^2}\nonumber\\
&=\frac{2e^2}{\hbar c}\int
\frac{d\theta k_F(\theta)}{4\pi^2}\frac{\left|v^*_{F}(\theta)\right|}{2}
\frac{2\Gamma^*_{\rm FS}}{\Omega^2+4\left(\Gamma^*_{\rm FS}\right)^2}
\end{align}
The second equality is found by symmetrizing $x$ and $y$ directions. Integrating as Eq.~\eqref{Kdef}, we  find the kinetic energy as
\begin{equation}
K_{\rm Drude}=\frac{2}{\pi}\int_0^\infty\left(\frac{\hbar c}{e^2}\right)\sigma_{\rm qp}(\Omega)d\Omega=\int\frac{d\theta k_F(\theta)}{4\pi^2}\left|v^*_{F}(\theta)\right|\label{KFLdef}
\end{equation}
This can be compared to the Drude weight of the computed optical conductivity.

The foregoing is general; in the three-band model of interest here the matrix ${\bf Z}={\rm diag}\left((1-\frac{\partial \Sigma}{\partial \omega})^{-1},1,1\right)$ while ${\bf\Delta_\mu}\sim {\rm diag}(-{\rm Re}\Sigma(\omega=0),0,0)$.
For the hole doping 0.24 data shown here, ${\rm Re}\Sigma=3.7$ eV, $(1-\frac{\partial \Sigma}{\partial \omega})^{-1}=0.16$. the Fermi surface implied by this ${\bf \Delta}_\mu$ has an area corresponding to a hole doping value of around 0.26. the slight deviation from the exact doping value 0.24 is due to the temperature/broadening effect.  Evaluating  Eq.~\eqref{KFLdef} yields $K_{\rm Drude}=0.07$ eV. the Drude weight of calculated conductivity (shown in Fig.~\ref{compWebercond0p24}) is 0.06 eV. The close agreement indicates that the low-frequency feature of our calculation is reliable.

We have also done the same calculation for 0.18 hole doping, in which case $\varepsilon_d=-5.4$ eV, $\varepsilon_p=-5.06$ eV, ${\bf \Delta}_\mu={\rm diag}(-3.6 {\rm eV},0,0)$, ${\bf Z}={\rm diag}(0.11,1,1)$. In this case the area enclosed by the Fermi surface indicates a hole doping of around 0.06, indicating that at the high temperatures that we study, the model is not yet in the Fermi liquid regime. We have found that $K=0.04$ eV from Eq.~\eqref{KFLdef}, comparing to the Drude weight of calculated conductivity (not shown) 0.033 eV. Despite the deviation from the Luttinger theorem the agreement is also good.

\end{document}